\documentclass[a4paper,12pt,oneside]{book}

\usepackage[utf8]{inputenc}
\usepackage{lmodern}
\usepackage{amsmath}
\usepackage{amsfonts}
\usepackage{braket}
\usepackage{anyfontsize}
\usepackage[toc,page]{appendix}
\usepackage{amssymb}
\usepackage[caption = false]{subfig}
\usepackage{graphicx}
\usepackage{subcaption}
\usepackage[left=2cm, right=2cm, top=2.54cm, bottom=2.54cm]{geometry}
\usepackage{float}
\usepackage{notoccite}
\usepackage{hyperref}
\usepackage{graphics}
\usepackage{wrapfig}
\usepackage{titlesec}
\usepackage{setspace}
\usepackage{acronym}   
\newpagestyle{mystyle}{\setfoot[\thepage][][]{}{}{\thepage}}
\titleformat
{\chapter} % command
[display] % shape
{\bfseries\Huge} % format
{Chapter \thechapter} % label
{0.3ex} % sep
{	\rule{\textwidth}{0.5pt}
	\vspace{-2.8ex}
	\centering
} % before-code
[
\vspace{-0.5ex}%
\rule{\textwidth}{0.5pt}
] % after-code
\titleformat{\section}[block]
{\bfseries\large}{\thesection}{1em}{}
\begin{document}
  \begin{titlepage}
  	\begin{center}  
  		\begin{spacing}{1.0}
   	{\fontsize{24pt}{28.8pt}\selectfont	\textbf{{\huge ENERGY CALCULATION OF PENTAQUARKS USING THOMAS FERMI QUARK MODEL: A THEORETICAL STUDY\\}}}
  		\end{spacing}		
  		\vspace{2.5cm}
  		{\fontsize{14pt}{16.8pt}\selectfont {\large \textbf{A Dissertation\\ \vspace{0.3cm}Submitted to the Central Department of Physics, Tribhuvan University, Kirtipur in the Partial Fulfillment for the\\ \vspace{0.2cm}Requirement of Master's Degree of Science in Physics }}}
\vspace{3cm}
         \begin{figure}[H]
         \centering
         \includegraphics[width=4.5cm, height=5cm]{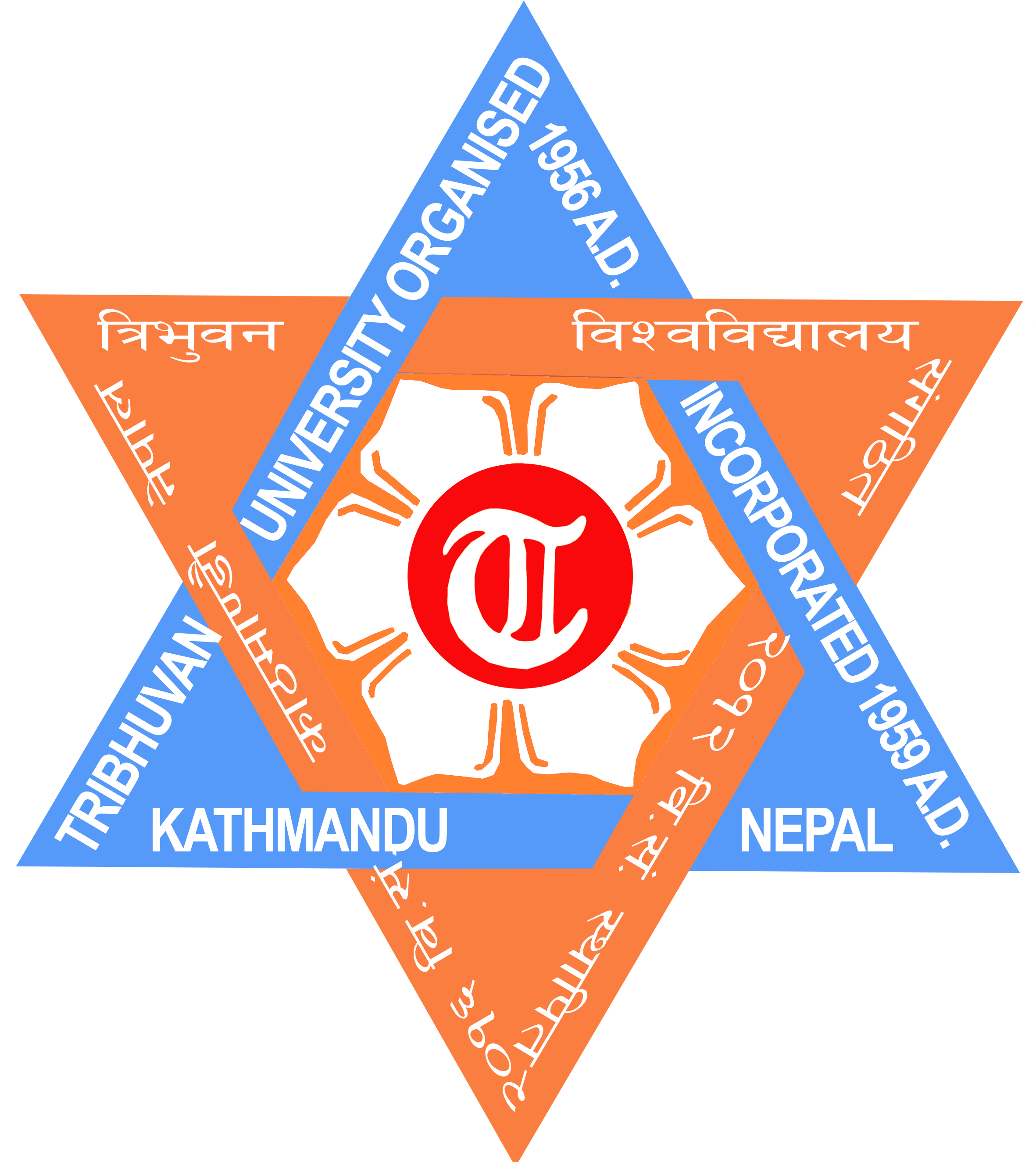}
         \end{figure}
		\vspace{0.8cm} 
		{\fontsize{24pt}{28.8pt}\selectfont\large{\textbf{By}}}\\
     	{\fontsize{20pt}{24pt}\selectfont\textbf{{\Large Bipin Aryal}}}\\{\fontsize{18pt}{22pt}\selectfont\textbf{{\Large Roll. No.: 3019/075}}}\\{\fontsize{28pt}{22pt}\selectfont\textbf{{\Large T.U. Reg. No.: 5-2-37-2024-2014}}}\\
     	{\fontsize{14pt}{16.8pt}\selectfont \textbf{September 2023}}\\%\vspace{0.1cm}Central Department of Physics\\ \vspace{0.1cm}Institute of Science and Technology\\ \vspace{0.1cm}
  		%Tribhuvan University,Kathmandu,Kirtipur\\ \vspace{0.2cm} 
  	  	\end{center}
  \end{titlepage}
\pagebreak
\frontmatter
	\vspace*{0.7cm}
	\begin{center}
		{\Large\textbf{RECOMMENDATION}}\\
	\end{center}
	\vspace{2em}
	\begin{spacing}{1.5}
	\addcontentsline{toc}{section}{Recommendation}
	\noindent It is certified that Mr.~Bipin~Aryal has carried out the dissertation work entitled  ``\textbf{ENERGY CALCULATION OF PENTAQUARKS USING THOMAS FERMI QUARK MODEL: A THEORETICAL STUDY}'' under our supervision and guidance.
	\par We recommend the dissertation in the partial fulfillment for the requirement of Master's Degree of Science in Physics.\\[2.5cm]
%\vspace*{2cm}
%\begin{figure}[H]
%\begin{minipage}{.5\textwidth}
%	\includegraphics[height=1in, width=2in]{sign_gck.png}
%\end{minipage}%
%\begin{minipage}{.5\textwidth}
%\centering
%	\includegraphics[height=1in, width=2in]{sign_sum.png}
%\end{minipage}
%\end{figure}
...................................  \hfill
...................................\\
\hspace{5cm}(Supervisor) \hfill  (Co-Supervisor) \hspace{1cm}\\
$\mathrm{{~Dr.~Gopi~Chandra~Kaphle}}$ \hfill $\ \mathrm{{~Dr.~Suman~ Baral}}$\\
\vspace{0.0001cm}Associate Professor\hfill Everest Institute of Science and
Technology\\
\vspace{0.0001cm}Central Department of Physics\hfill  Kathmandu, Nepal\\
\vspace{0.0001cm}Tribhuvan University, Kirtipur\hfill  \\
\vspace{0.0001cm}Kathmandu, Nepal\hfill  \\[3cm]
	\\
	\noindent Date: 10 September, 2023
	\end{spacing}
	\pagebreak
\begin{center}
	  	{\textbf{\Large ACKNOWLEDGEMENT}}\\
\end{center}
  	\begin{spacing}{1.5}
  		\addcontentsline{toc}{section}{Acknowledgement}
  	 I would like to take this opportunity to offer my sincere gratitude to Associate Professor Dr. Gopi Chandra Kaphle, who served as thesis supervisor, for his solid support and guidance during my research. My study technique and ideas have greatly benefited from his insightful observations, helpful advice, and encouraging feedback. I would also like to thank Dr. Suman Baral, my co-supervisor, for his significant contribution to my research and for providing me with valuable ideas that helped me to progress. Professor Dr. Walter Wilcox deserves special thanks as his remarkable ability to provide clarity in complex situations has guided me to the final result of this research.\\ \\
I would like to extend my appreciation to Mr. Mohan Giri for his exceptional teaching skills that made it easy for me to understand complex concepts. He was always available to answer my questions and provided excellent guidance, which has been a great help. I am also grateful to Mr. Nirmal Dangi for his invaluable contributions to productive discussions on various topics related to my research. His fresh perspective and willingness to help have been extremely beneficial.\\ \\
I would also like to thank my dear friend, Mr. Dipu Nepal, for his assistance, which has been invaluable in the successful completion of my research. We have had many fruitful discussions on various topics related to the research, and I could not have completed this work without his support.\\  \\
I am deeply grateful to my beloved family members for their unwavering encouragement, support, and patience as I went through my academic path. I am incredibly grateful for their existence in my life and their love and support have been a constant source of encouragement for me. \\ \\ 
Finally, I want to acknowledge the University Grant Commission for their partial financial support of this research. 
\end{spacing}
\pagebreak
\pagebreak
\begin{figure}[H]
	\centering
	\includegraphics[height=1.2in, width=1.2in]{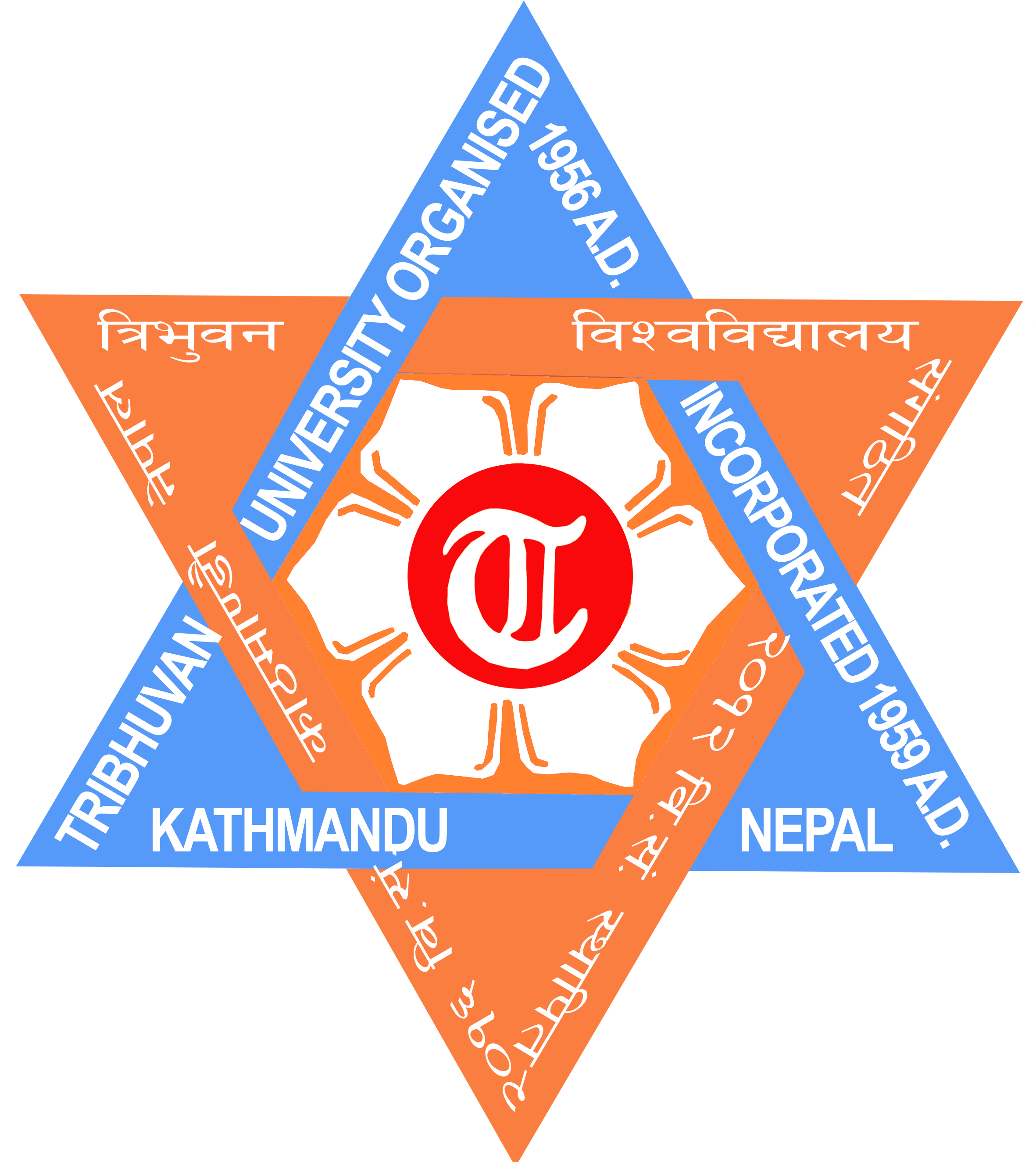}
\end{figure}
\vspace*{1.0cm}
\begin{center}
	{\Large\textbf{EVALUATION}}\\
\end{center}
	\addcontentsline{toc}{section}{Evaluation}
\begin{spacing}{1.5}
We certify that we have read this dissertation and in our opinion, it is satisfactory in scope and quality as a dissertation in partial fulfillment for the requirement of Master's Degree of Science in Physics.\\ 
\end{spacing}
\vspace{-2em}
\begin{center} \large\bf {\textbf {\underline{Evaluation Committee}}}
\end{center}
\vspace*{1.55cm}
$\overline{\mathrm{{Assoc.~Prof.~Dr.~Gopi~Chandra~Kaphle}}}$ \hfill $\overline{\mathrm{{Prof.~Dr.~Om~Prakash~Niraula}}}$\\
\hspace{5cm}(Supervisor) \hfill  (Head of Department) \hspace{1cm}\\
\vspace{0.0001cm}Central Department of Physics\hfill Central Department of Physics\\
\vspace{0.0001cm}Tribhuvan University, Kirtipur\hfill  Tribhuvan University, Kirtipur\\
\vspace{0.0001cm}Kathmandu, Nepal\hfill  Kathmandu, Nepal\\[2.20cm]
({$\overline{\mathrm{{External}~~{Examiner}}}$}) \hfill
($\overline{\mathrm{{Internal}~~{Examiner}}}$)\\[2cm]
Date: \underline{\,\,\,\,\,\,\,\,\,\,\,\,\,\,\,\,\,\,\,\,\,\,\,\,\,\,\,\,\,\,\,\,\,\,\,\,\,\,}
\pagebreak
\begin{center}
	{\Large\textbf{ABSTRACT}}\\
\end{center}
	\addcontentsline{toc}{section}{Abstract}
\begin{spacing}{1.5} 
The possible existence of a family of $uudc\overline{c}$ pentaquark system was explored through the application of Thomas Fermi Quark Model on the system. Considering a pocket of pentaquark contains $5$ quarks, the density functions, and energies of $10$ different pockets of pentaquarks were studied. Out of these $10$ different pockets, a single pocket of pentaquark was found to be more stable as its total energy was lowest compared to others. The kinetic, potential, and volume energies of all pockets of pentaquarks were calculated and the nature of the plots was investigated. The plots displayed an increasing trend for all energies except for kinetic energy, which exhibited a decreasing pattern.

\end{spacing}
\begin{center}
	\chapter*{{\huge\textbf{List of Abbreviations}}}
\end{center}
	\addcontentsline{toc}{section}{List of Abbreviations}
\begin{spacing}{1.5} 
\begin{tabular}{l c c c l }
   \textbf{CSM}& & & &Chiral Soliton Model \\
     \textbf{LHCb} & & & &Large Hadron Collider beauty\\
     \textbf{MIT} & & & &Massachusetts Institute of Technology \\ 
     \textbf{QCD} & & & &Quantum Chromodynamics \\
     \textbf{TF} & & & &Thomas Fermi \\ 
     \textbf{TFQM} & & & &Thomas Fermi Quark Model
\end{tabular}
\end{spacing}
\pagebreak
%\begin{center}
%	{\Large\textbf{List of Figures}}\\
%\end{center}
	\addcontentsline{toc}{section}{List of Figures}
%\begin{spacing}{1.5} 
\listoffigures
%\end{spacing}
\pagebreak
%\begin{center}
%	{\Large\textbf{List of Tables}}\\
%\end{center}
\addcontentsline{toc}{section}{List of Tables}
%\begin{spacing}{1.5} 
\listoftables
%\end{spacing}
\pagebreak
\tableofcontents
%Beginning of chapters:
\mainmatter
\chapter{Introduction}\label{chapter_1}
In this chapter, we give general information on some existing theories of particle physics and provide ideas based on which the research has been performed.
\section{General Consideration}
In 1964, Gell-Mann\cite{Gell-Mann} and Zweig\cite{Zweig} made a successful prediction about the existence of quarks as subatomic particles. They proposed that protons and neutrons, known as baryons, are composed of three quarks: two up quarks and one down quark for protons, and two down quarks and one up quark for neutrons\cite{neutrons}. This prediction set the stage for exploring multi-quark states, a quest that began before the development of Quantum Chromodynamics (QCD), the defining theory in this field. The advancement of QCD further fueled the investigation of multi-quark states, as it naturally predicted their existence based on the fundamental nature of QCD itself\cite{QCD itself}.\\ \\
In the early 21st century, numerous efforts were made to experimentally verify tetraquarks\cite{tetraquarks1, tetraquarks2} and pentaquarks\cite{pentaquarks1, pentaquarks2}, and various models were proposed to study these exotic particles and their properties. In 2015, the LHCb collaboration reported the discovery of two hidden-charm pentaquark states, named $P_{c}(4380)$ and $P_{c}(4450)$, in the $J/\psi p$ invariant mass spectrum of $\Lambda^{0}_{b} \longrightarrow J / \psi K^{-} p$\cite{pentaquarks1}. The masses and widths of these states, determined through a fit using Brieit-Wigner amplitudes, were found to be $M_{P_{c}(4380)}$ = $(4380 \pm 8 \pm 29)$ MeV, $\Gamma_{P_{c}(4380)}$ = $(205 \pm 18 \pm 86) $MeV, $M_{P_{c}(4450)}$ = $(4449.8 \pm 1.7 \pm 2.5)$ MeV, and $\Gamma_{P_{c}(4450)}$ = $(39 \pm 5 \pm 19)$ MeV\cite{reaction_masses}. Four years later, the LHCb Collaboration updated their results, proposing a new state called $P_{c}(4312)$, and splitting the previously observed $P_{c}(4450)$ into $P_{c}(4440)$ and $P_{c}(4457)$ states\cite{pentaquarks2}. \\ \\
The discovery of these ($P_{c}^{+}$) pentaquarks has generated significant interest in the theoretical investigation of pentaquarks. Various explanations have been suggested, including the chiral soliton model (CSM)\cite{CSM model}, the diquark model\cite{diquarkmodel1, diquarkmodel2, diquarkmodel3, diquarkmodel4}, the QCD sum rule approach\cite{QCDsumruleapproach1, QCDsumruleapproach2}, the MIT bag model\cite{MITbagmodel1, MITbagmodel2}, TF quark model\cite{TFquarkmodel}, and others. While QCD serves as the fundamental theory of strong interaction, studying the structure of hadrons and hadron-hadron interactions directly is challenging due to the non-perturbative nature of QCD in the low-energy regime\cite{lowenergyregime}. The proposed theoretical quark models overcome these limitations and provide insight into regions where QCD calculations are inaccessible.

\section{Thomas Fermi Quark Model}
The TF quark model (TFQM), developed by Dr. Walter Wilcox, incorporates the fundamental concepts of Thomas Fermi's atomic model\cite{TFatomicmodel1, TFatomicmodel2}. Wilcox discussed three different baryonic applications of the model: an equal mass nonrelativistic model with and without volume pressure, an ultra-relativistic limit constrained by volume pressure, and a color-flavor locking mass-less model\cite{TFquarkmodel}. Q. Liu and Walter Wilcox conducted numerical investigations of the non-relativistic aspects of the TF quark model and proposed concepts for incorporating spin splitting into the model\cite{QLiu}. Suman Baral and Walter Wilcox explored the application of TFQM to determine the potential existence of mesons with two or more quark-antiquark pairs\cite{SumanBaral}. Mohan Giri and colleagues utilized the TFQ Model to examine the patterns of quark distribution across a family of pentaquarks, $uudc\Bar{c}$\cite{MohanGiri}.
\section{Scope of Present work}
Mohan Giri $et.al$ in their research found that at least three TF functions with three different radii are required to study the distribution of quarks inside pentaquarks. They observed that for a pentaquark, heavy charm antiquark is limited to the innermost region. As the quark content increases, however, the heavy charm, rather than the anti-charm, is limited to the center region. In this research, I have extended the work done by Giri and his colleagues to study the system energy of those pentaquarks and inspect the energy trends to find out if any families of pentaquarks are stable.\\\\
The thesis is organized as follows. Chapter 2 provides a detailed description of the theoretical background necessary to address the problem at hand, including the Thomas Fermi Quark equations and general expressions for kinetic, potential, and total energies. Chapter 3 presents the methodology used to obtain our results, including consistency conditions and energy expressions for two different Cases. This chapter also includes a brief introduction to the Mathematica program used for calculations and generating plots. In Chapter 4, we present our results and interpret the associated plots. Chapter 5 includes a summary of our findings and suggestions for future research in the same field. Chapter 6 and Chapter 7 are the appendix sections that contain equations used in deriving the energy expressions for Case I and Case II respectively. Finally, Chapter 8 contains plots of density functions for $\eta \ge 3$. 
\chapter{Theoretical Background}
\label{chaper_2}
In this chapter, we establish the theoretical foundation for our research by grounding the Thomas Fermi Quark Model as a statistical model and the TF quark equations. We then obtain general expressions for the kinetic energy, potential energy, and total energy of the pentaquark family. 
  \section{Thomas Fermi Quark Model}\label{te}
In this section, we first interpret the TF quark model as a statistical model and show how kinetic and potential energies are expressed as a function of the density of states. The Thomas Fermi Quark equations are then derived by using the weighted interaction probabilities between constituents of pockets of pentaquarks. 
\subsection{Thomas Fermi Statistical Model}
Thomas-Fermi Statistical Model is a semi-classical model in the sense that it incorporates features of both classical and quantum mechanics. It treats particles as a Fermi gas at an absolute zero temperature. Although it utilizes Fermi statistics for the distribution of particles it does not have a quantum mechanical wave function, but, rather, a central function related to particle density. In the Thomas-Fermi Quark Model, quarks are thought to be evenly distributed across each volume element $\Delta$V. However, the quark density $n_q(r)$ differs from one volume element to another. For small volume element $\Delta$V, quarks can fill a volume in spherical momentum space $V_F$ up to the Fermi momentum $p_F$. The spherical momentum space volume $V_F$ can be expressed as follows
\begin{equation}
    V_F=\frac{4}{3}\pi p_F^3(\vec{r}).
\end{equation}
The phase space volume represents the volume in momentum space that is occupied by quarks within the small volume element $\Delta$V at position $\vec{r}$.
\begin{equation}
    \Delta V_{ph}= V_F \Delta V = \frac{4}{3}\pi p_F^3(\vec{r}) \Delta V
\end{equation}
The number of quarks within $\Delta V_{ph}$ is given by,
\begin{equation}
    \Delta N_{ph}=\frac{g_I}{h^3}\Delta V_{ph}=\frac{4 \pi g_I}{3h^3}p_F^3(\vec{r})\Delta V
    \label{ph} 
\end{equation}
$g_{I}$ is the degeneracy of quark with flavor $I$ . $h$ is Planck's constant, and $p_{F}(\vec{r})$ is the Fermi momentum at position $\vec{r}$.\\ \\
The number of quarks within a volume element $\Delta V$ is
\begin{equation}
    \Delta N=n_q(\vec{r}) \Delta V 
    \label{hp}
\end{equation}
where, $n_q(\vec{r})$ is the quark density. Combining equations (\ref{hp}) and (\ref{ph}), we obtain
\begin{equation}
    n_q(\vec{r}) =\frac{4 \pi g_I}{3h^3}p_F^3(\vec{r})
    \label{hppp}
\end{equation}
The distribution function $F_{(\vec{r})}(p)dp$ for quarks with momentum $p$ at position $\vec{r}$ is 
\begin{equation}
F_{(\vec{r})}(p)dp=\left\{\begin{array}{lr}
        \frac{4 \pi p^2 dp}{\frac{4}{3}\pi p_F^3(\vec{r})} &\quad\text{if $p<p_f(\vec{r})$}\\
        0 &\quad\text{otherwise} \\ 
        \end{array} \right.
    \end{equation}
This expression provides a way to describe the distribution of quarks in momentum space at a specific position $\vec{r}$, taking into account the Fermi momentum $p_{F}(\vec{r})$ as a dividing point for the distribution.\\ \\
The kinetic energy at position $\vec{r}$ can be obtained by integrating over the momentum space from $0$ to $p_{F}$
\begin{equation}
    t(\vec{r})=\int^{p_f}_{0}\frac{p^2}{2m_I}n_q(\vec{r}) F_{(\vec{r})}(p)dp=C_F\left(n_q(\vec{r}) \right)^\frac{5}{3}
\label{t}
\end{equation}
Where,
\begin{equation}
    C_F= \frac{3 \left( 6 \pi^2 \hbar^3 \right)^\frac{5}{3}}{20 \pi^2 \hbar^3 m_I (g_I)^\frac{2}{3}}
    \label{eq6}
\end{equation}
For total energy, the integral is taken over the three-dimensional space,$\vec{r}$, to consider the contributions of kinetic energy at all positions in the system. 
\begin{equation}
    T=\int t(\vec{r})d^3r
    \label{eq7}
\end{equation}
Therefore, from equations (\ref{t}), (\ref{eq6}) and (\ref{eq7}), the kinetic energy is given by
\begin{equation}
    T=\int \frac{3 \left( 6 \pi^2 \hbar^3 \right)^\frac{5}{3}}{20 \pi^2 \hbar^3 m_I (g_I)^\frac{2}{3}} \left(n_q(\vec{r}) \right)^\frac{5}{3} d^3r
    \label{kine}
\end{equation}
The total potential energy of the system is 
\begin{equation}
    U=V_N\int \int \frac{n_q(\vec{r}) n_q(\vec{r}')}{|\vec{r}-\vec{r'}|}d^3r d^3r' 
    \label{poten}
\end{equation}
$V_{N}$ is a normalization constant that depends on the interaction between quarks. Therefore, the total energy of the system of quarks is
\begin{equation}
    E=C_F\int \left(n_q(\vec{r}) \right)^\frac{5}{3}d^3r+V_N\int \int \frac{n_q(\vec{r}) n_q(\vec{r}')}{|\vec{r}-\vec{r}'|}d^3r d^3r'
    \label{E}
\end{equation}
Lagrange multiplier term $\lambda(\int n_q(\vec{r})  d^3r-N_q)$ is added to equation (\ref{E}) to minimize energy keeping total number of quarks $N_q$ constant.\\
   \subsection{Thomas Fermi Quark Equations}\label{tf} 
The two TF quark equations for a system of pentaquark are\cite{MohanGiri} 
   \begin{equation}
    \overline{\alpha}_I \frac{d^2\overline{f_I(x)}}{dx^2}= -\frac{6\eta}{5(5\eta-1)} \sum_J g_J \frac{(f_J(x))^\frac{3}{2}}{\sqrt{x}}
    \label{700a}
    \end{equation}
\begin{equation}
\alpha_I \frac{d^2 f_I(x)}{dx^2}=-\frac{6\eta}{5(5\eta-1)}\sum_I \overline{g}_I \frac{\left( \overline{f_I(x)} \right)^\frac{3}{2}}{\sqrt{x}} - \frac{9\eta}{10(5\eta-1)}\left[\frac{(N_I g_I-1)}{N_I} \frac{\left( f_I(x) \right)^\frac{3}{2}}{\sqrt{x}} +\sum_{I \ne J} g_J \frac{\left( f_J(x) \right)^\frac{3}{2}}{\sqrt{x}} \right].
\label{700b}
\end{equation}
These equations are used to build consistency equations of various functions that exist inside different regions of the pentaquark system. These are also used in forming system energies of the pentaquark system.
\section{Expression for Total Energy}
The total energy in the Thomas Fermi Quark Model is the sum of kinetic, potential, and volume energy. The expressions for kinetic and potential energy are given below. The volume energy solely depends on the outermost boundary of the given pentaquark system.  \\ \\ 
The general expression for kinetic energy obtained from the Thomas Fermi Model in terms of quark densities is\cite{MohanGiri} 
\begin{equation}
T=\frac{12}{5\pi} \frac{\frac{4}{3}\alpha_s \frac{4}{3}g^2}{a}\left(\frac{3 \pi \eta}{2} \right)^\frac{1}{3}\left[\displaystyle \sum_{I} \alpha^I g^I S^I +\displaystyle \sum_{I} \overline{\alpha}^I \overline{g}^I \overline{S}^I \right]
\label{T}
\end{equation}
This equation has two integrals $S^I$ and  $\overline{S}^I$ which depend upon the quark densities and radius of the boundaries of the quarks and they are the sole factors of kinetic energy for a particular pentaquark (i.e. for a fixed value of $\eta$). \\ \\ 
The general expression for potential energy as given by Thomas Fermi Quark Model is\cite{MohanGiri}
\begin{equation}
U=-\frac{18}{25 \pi}\frac{\eta}{\left(\eta-\frac{1}{5}\right)}\frac{\frac{4}{3}\alpha_s \frac{4}{3}g^2}{a}\left(\frac{3 \pi \eta}{2} \right)^\frac{1}{3} \left[\displaystyle \sum_I \frac{(g_I N_I-1)g_I}{N_I}K_{II}+\sum_{I \ne J} g_I g_J K_{IJ} + \frac{4}{3}\sum_{I, J} \overline{g}_I g_J K_{\overline{I}J}\right]
\label{U}
\end{equation}
$K_{II}$, $K_{IJ}$ \&  $K_{\overline{I}J}$ are integrals that can be found out using consistency conditions discussed in chapter \ref{chapter_3}. $g_I$ and $N_I$ respectively are the degeneracy and the number of quarks at that degeneracy for a particular flavor $I$. Their values, at ground state, are fixed for a particular pentaquark. 
\chapter{Methodology}
\label{chapter_3}
In this chapter, we discuss consistency conditions, system energies, and the parameters that are used to obtain results that fulfill our objectives. The computational details involved in the execution of the program is discussed in the last section of the chapter.  
\section{Pocket of Pentaquarks, $\eta$}
In our study of pentaquarks, we deal with the pockets of pentaquarks rather than a single pentaquark. A single pocket of pentaquark consists of 5 quarks. With the increase in the number of pockets of pentaquarks the quark number increases in the multiple of 5. The family stability of 10 different pockets of the pentaquark system, $uudc\overline{c}$, are investigated in this research. 
\section{Case I: $\eta =1$}
In Case I, all quarks are distributed in the inner region, charm and light quarks in the middle region, and the light quarks in the outermost region. It is present in only a single pocket of the pentaquark system\cite{MohanGiri}.
    \begin{figure}[h]
     \centering
\includegraphics[width=0.49\linewidth]{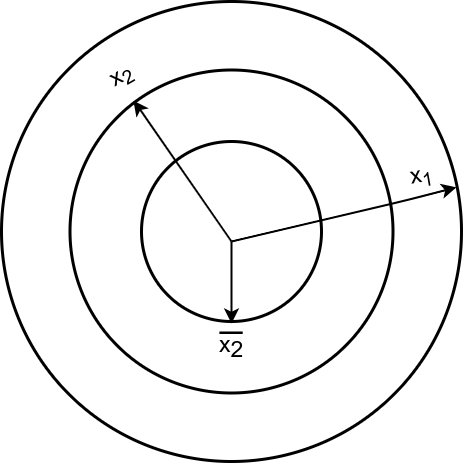} 
\caption{Distribution of quarks for the Case of $\eta = 1$}
\label{fig:1}
\end{figure}
As shown in fig \ref{fig:1}, in the innermost region, $0<x<\overline{x_2}$, all three quarks are present. In the middle region, $\overline{x_{2}}<x< x_2$, charm quarks and light quarks are present whereas in the outermost region, $x_{2}<x< x_1$, only light quarks are present.\\    \\
\subsection{Consistency Conditions}
We have two Thomas Fermi equations as
\begin{equation}
    \overline{\alpha}_I \frac{d^2\overline{f_I(x)}}{dx^2}= -\frac{6\eta}{5(5\eta-1)} \sum_J g_J \frac{(f_J(x))^\frac{3}{2}}{\sqrt{x}}
    \label{1}
    \end{equation} 
   \begin{equation}
    \alpha_I \frac{d^2 f_I(x)}{dx^2}=-\frac{6\eta}{5(5\eta-1)}\sum_I \overline{g}_I \frac{\left( \overline{f_I(x)} \right)^\frac{3}{2}}{\sqrt{x}} - \frac{9\eta}{10(5\eta-1)}\left[\frac{(N_I g_I-1)}{N_I} \frac{\left( f_I(x) \right)^\frac{3}{2}}{\sqrt{x}}
     +\sum_{I \ne J} g_J \frac{\left( f_J(x) \right)^\frac{3}{2}}{\sqrt{x}} \right]
\label{2}
      \end{equation}
      \noindent Above two equations can be expanded with three different functions namely $f_1(x)$ for light quarks $(uud)$, $f_2(x)$ for heavy quark $c$ \& $\overline{f_2(x)}$ for heavy anti-quark $\overline{c}$ as 
      \begin{equation}
           \overline{\alpha}_2 \frac{d^2\overline{f_2(x)}}{dx^2}= -\frac{6\eta}{5(5\eta-1)} \biggl\{ g_1 \frac{(f_1(x))^\frac{3}{2}}{\sqrt{x}} + g_2 \frac{(f_2(x))^\frac{3}{2}}{\sqrt{x}} \biggr\}
           \label{yes}
      \end{equation}
   \begin{equation}
    \alpha_1 \frac{d^2 f_1(x)}{dx^2}= -\frac{6 \eta}{5(5 \eta -1)}\overline{g_{2}} \frac{(\overline{f_{2}(x)})^\frac{3}{2}}{\sqrt{x}}-\frac{9\eta}{10(5\eta-1)}\left[\frac{(N_1 g_1-1)}{N_1} \frac{\left( f_1(x) \right)^\frac{3}{2}}{\sqrt{x}}
     + g_2 \frac{\left( f_2(x) \right)^\frac{3}{2}}{\sqrt{x}} \right]
\label{no}
      \end{equation}
      \begin{equation}
    \alpha_2 \frac{d^2 f_2(x)}{dx^2}=-\frac{6\eta}{5(5\eta-1)}    \overline{g_2} \frac{\left( \overline{f_2(x)} \right)^\frac{3}{2}}{\sqrt{x}} - \frac{9\eta}{10(5\eta-1)}\left[\frac{(N_2 g_2-1)}{N_2} \frac{\left( f_2(x) \right)^\frac{3}{2}}{\sqrt{x}}
     + g_1 \frac{\left( f_1(x) \right)^\frac{3}{2}}{\sqrt{x}} \right]
\label{dilemma}
      \end{equation}
These three second-order differential equations are responsible for generating consistency conditions in different regions of pentaquarks. 
       \begin{itemize}
        \item \textbf{In region $0<x<\overline{x_2}$} \\ 
          \end{itemize}
          \vspace{-1cm}
          \noindent In this region, all three functions are present but we use the linear relationship of other functions with $\overline{f_2(x)}$ as 
          \begin{equation}
          \begin{split}
              f_1(x) = \beta_{1\overline{2}} \overline{f_2(x)}\\            
              f_2(x) = \beta_{2\overline{2}} \overline{f_2(x)}
              \end{split}
              \label{11}
          \end{equation}
          \hfill where, $\beta_{1\overline{2}}$ \& $\beta_{2\overline{2}}$ are proportionality constants.
          \\\\
         Equation \ref{yes} gives 
       \begin{equation}
       \frac{d^2\overline{f_2(x)}}{dx^2} = -\frac{6\eta}{5(5\eta-1) \overline{\alpha}_2} \biggl\{ g_1 (\beta_{1\overline{2}})^\frac{3}{2} + g_2 (\beta_{2\overline{2}})^\frac{3}{2}
        \biggr\}\frac{\overline{f_2(x)}^\frac{3}{2}}{\sqrt{x}}
        \label{12}
      \end{equation}
      \noindent Equation \ref{no} gives 
      \begin{equation}
      \frac{d^2 \overline{f_2(x)}}{dx^2}= \left[- \frac{6 \eta}{5 (5 \eta -1) \beta_{1\overline{2}}}\overline{g_{2}} -\frac{9\eta}{10(5\eta-1)\beta_{1\overline{2}}}\left\{\frac{(N_1 g_1-1)(\beta_{1\overline{2}})^\frac{3}{2}}{N_1}
     + g_2 (\beta_{2\overline{2}})^\frac{3}{2} \right\}\right]\frac{\left( \overline{f_2(x) }\right)^\frac{3}{2}}{\sqrt{x}}
     \label{13}
      \end{equation}
      \noindent Equation \ref{dilemma} gives
      \begin{equation}
      \frac{d^2 \overline{f_2(x)}}{dx^2}= \left[-\frac{6\eta}{5(5\eta-1) \alpha_2 \beta_{2\overline{2}}}    \overline{g_2} - \frac{9\eta}{10(5\eta-1) \alpha_2 \beta_{2\overline{2}}}\left\{\frac{(N_2 g_2-1)(\beta_{2\overline{2}})^\frac{3}{2}}{N_2}      + g_1 (\beta_{1\overline{2}})^\frac{3}{2}\right\}\right]\frac{\left( \overline{f_2(x)} \right)^\frac{3}{2}}{\sqrt{x}} 
   \label{14}
      \end{equation}
      \noindent In a single form eqns \ref{12}, \ref{13}, \ref{14} can be written as 
      \begin{equation}
       \frac{d^2\overline{f_2(x)}}{dx^2} = \overline{Q_2} \frac{\left(\overline{f_2(x)}\right)^\frac{3}{2}}{\sqrt{x}}
          \label{15}
      \end{equation}
      where,  
      \begin{equation}
            \overline{Q_2} 
          \begin{aligned} &
               = -\frac{6\eta}{5(5\eta-1) \overline{\alpha}_2} \biggl\{ g_1 (\beta_{1\overline{2}})^\frac{3}{2} + g_2 (\beta_{2\overline{2}})^\frac{3}{2}
        \biggr\}\\ &
       = - \frac{6 \eta}{5 (5 \eta -1) \beta_{1\overline{2}}}\overline{g_{2}} -\frac{9\eta}{10(5\eta-1)\beta_{1\overline{2}}}\left[\frac{(N_1 g_1-1)(\beta_{1\overline{2}})^\frac{3}{2}}{N_1} 
     + g_2 (\beta_{2\overline{2}})^\frac{3}{2} \right]\\ &
     = 
     -\frac{6\eta}{5(5\eta-1) \alpha_2 \beta_{2\overline{2}}}    \overline{g_2} - \frac{9\eta}{10(5\eta-1) \alpha_2 \beta_{2\overline{2}}}\left\{\frac{(N_2 g_2-1)(\beta_{2\overline{2}})^\frac{3}{2}}{N_2}      + g_1 (\beta_{1\overline{2}})^\frac{3}{2}\right\}
          \end{aligned}
          \label{16}
      \end{equation}\\
This second-order differential equation establishes the relationship between the anti-charm function, $\overline{f_{2}(x)}$, and the other functions present inside the region. 
      \begin{itemize}
          \item \textbf{In region $\overline{x_2}<x< x_2$}
      \end{itemize}
      In this region, two functions $f_2(x)$ \& $f_1(x)$ are present and we use linear relationship of $f_1(x)$ with $f_2(x)$ as
      \begin{equation}
          f_1(x) = \gamma_{12} f_2(x)
          \label{17}
      \end{equation}
       \noindent Equation \ref{yes} gives           
       \begin{equation}
      \frac{(f_2(x))^\frac{3}{2}}{\sqrt{x}} = 0
      \label{18}
      \end{equation}
      This leads to meaningless results. So we drop this equation. \\
    \noindent Equation \ref{no} gives 
      \begin{equation}
      \frac{d^2 f_2(x)}{dx^2}= -\frac{9\eta}{10(5\eta-1)  \gamma_{12} }\left\{\frac{(N_1 g_1-1)(\gamma_{12})^\frac{3}{2}}{N_1} + g_2 \right\}\frac{\left( f_2(x) \right)^\frac{3}{2}}{\sqrt{x}} 
  \label{19}
      \end{equation}
  Equation \ref{dilemma} gives
      \begin{equation}
     \frac{d^2 f_2(x)}{dx^2}= - \frac{9\eta}{10(5\eta-1)\alpha_2}\left\{\frac{(N_2 g_2-1)}{N_2} 
     + g_1 (\gamma_{12})^\frac{3}{2} \right\}\frac{\left( f_2(x) \right)^\frac{3}{2}}{\sqrt{x}}
     \label{20}
      \end{equation}
      \noindent In a single form eqns \ref{19}, \ref{20} can be written as 
      \begin{equation}
         \frac{d^2f_2(x)}{dx^2} = Q_2 \frac{\left(f_2(x)\right)^\frac{3}{2}}{\sqrt{x}}
          \label{21}
      \end{equation}
      where, 
      \begin{equation}
      \begin{aligned} Q_2    &
        = -\frac{9\eta}{10(5\eta-1)  \gamma_{12} }\left\{\frac{(N_1 g_1-1)(\gamma_{12})^\frac{3}{2}}{N_1} + g_2 \right\}\\ &
                    = -\frac{9\eta}{10(5\eta-1)\alpha_2}\left\{\frac{(N_2 g_2-1)}{N_2} 
     + g_1 (\gamma_{12})^\frac{3}{2} \right\}
          \end{aligned}
          \label{22}
      \end{equation}\\
         This second-order differential equation defines the link between the charm function, $f_{2}(x)$, and the other function present inside the region. 
    \begin{itemize}
   \item \textbf{In region $x_2<x< x_1$}
    \end{itemize}
    \noindent In this region, $f_2(x) = 0$ \& $\overline{f_2(x)}=0$. \\
    \noindent Equation \ref{yes} gives 
    \begin{equation}
           0 = -\frac{6\eta}{5(5\eta-1)} \biggl\{ g_1 \frac{(f_1(x))^\frac{3}{2}}{\sqrt{x}} \biggr\}
           \label{dipu}
    \end{equation}
           This leads to meaningless result. So we drop this equation. \\
    \noindent Equation \ref{no} gives 
   \begin{equation}
        \frac{d^2 f_1(x)}{dx^2}= -\frac{9\eta}{10(5\eta-1)}\left[\frac{(N_1 g_1-1)}{N_1} \frac{\left( f_1(x) \right)^\frac{3}{2}}{\sqrt{x}}\right]
        \label{bikash}
      \end{equation}
            \noindent Equation \ref{dilemma} gives 
      \begin{equation}
0 = - \frac{9\eta}{10(5\eta-1)}\left[g_1 \frac{\left( f_1(x) \right)^\frac{3}{2}}{\sqrt{x}} \right]
\label{bimal}
      \end{equation}
         This leads to meaningless results. So we drop this equation. \\
       In a single form eqn \ref{bikash} can be written as 
      \begin{equation}
    \frac{d^2 f_1(x)}{dx^2}= Q_1 \frac{\left( f_1(x) \right)^\frac{3}{2}}{\sqrt{x}}
    \label{9}
      \end{equation}
     \begin{equation}\text{where, }Q_1 = \frac{-9\eta}{10(5\eta-1)}\frac{(N_1g_1 - 1)}{N_1}
     \label{10}
       \end{equation}
This second-order differential equation defines the existence of light quarks in the outermost region of the pentaquark system. \\ \\
 \noindent Therefore, eqns \ref{9}, \ref{15}, \ref{21} are the required consistency conditions with the values of $Q_1$, $\overline{Q_2}$ \& $Q_2$ in eqns \ref{10}, \ref{16} \& \ref{22} respectively. 
        \subsection{Expression of Kinetic Energy}
        The expression of kinetic energy given in eqn \ref{T} can be written in quark densities and radius of boundaries as 
        \begin{equation}
    T = \sum_I \int^{r_{max}} d^3r \frac{3(6\pi^2\hbar^3)^\frac{5}{3}}{20\pi^2\hbar^3m^I(g^I)^\frac{2}{3}} \bigl(n^I(r)\bigr)^\frac{5}{3} + \sum_I \int^{r_{max}} d^3r \frac{3(6\pi^2\hbar^3)^\frac{5}{3}}{20\pi^2\hbar^3\overline{m}^I (\overline{g^I})^\frac{2}{3}} \bigl(\overline{n^I(r)}\bigr)^\frac{5}{3}
    \label{23}
\end{equation}
  After solving eqn \ref{23} and careful reduction, the final expression of kinetic energy for the Case I looks like
        \begin{equation}
    \begin{split}
        \frac{(K.E)}{C} &= \frac{4}{7} \biggl[\alpha_1g_1\bigl\{(\beta_{1\overline{2}})^\frac{5}{2} - (\gamma_{12})^\frac{5}{2} \left(\frac{\beta_{1\overline{2}}}{\gamma_{12}}\right)^\frac{5}{2} \bigr\} +\left. \overline{\alpha_2}\overline{g_2}\biggr] \sqrt{x} (\overline{f_2(x)})^\frac{5}{2} \right|_{\overline{x_2}}  + \left. \frac{4}{7} \alpha_2g_2\sqrt{x} (f_2(x))^\frac{5}{2} \right|_{x_2} \\ & \quad +\left. \frac{4}{7} \alpha_1g_1 \sqrt{x} (f_1(x))^\frac{5}{2}\right|_{x_1}  + \frac{5\overline{N_2}}{21\eta} \biggl[\left(\frac{\beta_{1\overline{2}}}{\gamma_{12}}\right)^2\frac{\overline{Q_2}}{Q_2}\bigl\{\alpha_1g_1(\gamma_{12})^\frac{5}{2} + \alpha_2g_2\bigr\} - \alpha_1g_1 (\beta_{1\overline{2}})^\frac{5}{2} -\alpha_2g_2 (\beta_{2\overline{2}})^\frac{5}{2} \\ & \quad -\left. \overline{\alpha_2}\overline{g_2}\biggr]\frac{d\overline{f_2(x)}}{dx}\right|_{\overline{x_2}} + \frac{5}{21\eta} \biggl[\frac{(\gamma_{12})^2 \alpha_{1}g_{1}}{Q_1}  \biggl\{N_2Q_2  +\left(\frac{\beta_{1\overline{2}}}{\gamma_{12}}\right)\overline{N_2}\overline{Q_2}- (\beta_{2\overline{2}})^\frac{3}{2} \overline{N_2}Q_2\biggr\}-\biggl\{\alpha_1g_1(\gamma_{12})^\frac{5}{2} \\ & \quad +\left. \alpha_2g_2\biggr\}\biggl\{N_2+\left(\frac{\beta_{1\overline{2}}}{\gamma_{12}}\right)\frac{\overline{N_2}\overline{Q_2}}{Q_2} - (\beta_{2\overline{2}})^\frac{3}{2}\overline{N_2}\biggr\}\biggr]\frac{df_2(x)}{dx}\right|_{x_2}  - \frac{5\alpha_1g_1}{21\eta}\biggl[ N_1 + \gamma_{12}\frac{N_2Q_2}{Q_1} \\ & \quad + (\gamma_{12})^\frac{3}{2} \biggl\{(\beta_{2\overline{2}})^\frac{3}{2}\overline{N_2} - N_2\biggr\}- (\beta_{1\overline{2}})^\frac{3}{2}\overline{N_2} - (\gamma_{12})(\beta_{2\overline{2}})^\frac{3}{2}\frac{\overline{N_2}Q_2}{Q_1}  + \left. (\gamma_{12})\left(\frac{\beta_{1\overline{2}}}{\gamma_{12}}\right)\frac{\overline{N_2}\overline{Q_2}}{Q_1}\biggr]\frac{df_1(x)}{dx}\right|_{x_1}
        \end{split}
        \label{32}
    \end{equation}
    The appendices used to evaluate the final expression of kinetic energy for Case I are given in \autoref{appendix_I}.
      \subsection{Expression of Potential Energy}
        The expression of potential energy given in eqn \ref{U} can be deduced to
\begin{equation}
   - \frac{U}{C} = \frac{(g_1 N_1 - 1)}{N_1} g_1 K_{11} + \frac{(g_2 N_2 - 1)}{N_2} g_2 K_{22} + \frac{(\overline{g_2} \overline{N_2} - 1)}{\overline{N_2}} \overline{g_2} K_{\overline{2}\overline{2}} + g_{1} g_{2} K_{12} + g_{2} g_{1} K_{21} + \tfrac{4}{3} \overline{g_2} g_1 K_{\overline{2}1} + \tfrac{4}{3} \overline{g_2} g_2 K_{\overline{2}2}
  \label{41}
\end{equation}
The seven different constants on which the potential energy of Case I depends are
\begin{equation}
      \begin{split}
         K_{11} & = \frac{4}{7} \biggl[ \frac{ \left(\frac{\beta_{1\overline{2}}}{\gamma_{12}}\right)^\frac{5}{2} (\gamma_{12})^3}{Q_2} - \left. \frac{(\beta_{1\overline{2}})^3}{\overline{Q_2}} \biggr] \sqrt{x} (\overline{f_2(x)})^\frac{5}{2} \right|_{\overline{x_2}} + \left. \frac{4}{7} \biggl[\frac{(\gamma_{12})^\frac{5}{2}}{Q_1} - \frac{(\gamma_{12})^3}{Q_2} \biggr] \sqrt{x} (f_2(x))^\frac{5}{2} \right|_{x_2} \\ & \quad -\left. \frac{4}{7 Q_1} \sqrt{x} (f_1(x))^\frac{5}{2} \right|_{x_1}  + \Biggl[ \frac{\overline{N_2}}{3 \eta} \Biggl\{ \frac{(\beta_{1\overline{2}})^3}{\overline{Q_2}}  - \frac{(\beta_{1\overline{2}})^\frac{3}{2} (\gamma_{12})^\frac{3}{2}  \left(\frac{\beta_{1\overline{2}}}{\gamma_{12}}\right)}{Q_2} + \frac{\overline{Q_2}}{{Q_2}^2} (\gamma_{12})^3  \left(\frac{\beta_{1\overline{2}}}{\gamma_{12}}\right)^2 \\ & \quad - \frac{(\beta_{1\overline{2}})^\frac{3}{2} (\gamma_{12})^\frac{3}{2}  \left(\frac{\beta_{1\overline{2}}}{\gamma_{12}}\right)}{Q_2} \Biggr\} + \frac{5 \overline{N_2}}{21 \eta} \Biggl\{ \frac{(\beta_{1\overline{2}})^3}{\overline{Q_2}} - \left. \frac{\overline{Q_2}}{{Q_2}^2}  \left(\frac{\beta_{1\overline{2}}}{\gamma_{12}}\right)^2 (\gamma_{12})^3 \Biggr\} \Biggr] \frac{d\overline{f_2(x)}}{dx} \right|_{\overline{x_2}}  + \Biggl[ \frac{\overline{N_2}}{3 \eta} \Biggl\{ \frac{(\beta_{1\overline{2}})^\frac{3}{2} (\gamma_{12})^\frac{3}{2}}{Q_2} \\ & \quad - \frac{(\beta_{1\overline{2}})^\frac{3}{2} \gamma_{12}}{Q_1} - \frac{(\beta_{2\overline{2}})^\frac{3}{2}(\gamma_{12})^3}{Q_2} + \frac{(\beta_{2\overline{2}})^\frac{3}{2} (\gamma_{12})^\frac{5}{2}}{Q_1}  + \frac{(\beta_{1\overline{2}})^\frac{3}{2} (\gamma_{12})^\frac{3}{2}}{Q_2} - \frac{\overline{Q_2}}{{Q_2}^2} (\gamma_{12})^3  \left(\frac{\beta_{1\overline{2}}}{\gamma_{12}}\right) - \frac{(\beta_{1\overline{2}})^\frac{3}{2} \gamma_{12}}{Q_1} \\ & \quad + \frac{(\gamma_{12})^\frac{5}{2} (\beta_{2\overline{2}})^\frac{3}{2}}{Q_1} + \frac{\overline{Q_2}}{{Q_1}^2} (\gamma_{12})^2  \left(\frac{\beta_{1\overline{2}}}{\gamma_{12}}\right)   - \frac{Q_2}{{Q_1}^2} (\beta_{2\overline{2}})^\frac{3}{2} (\gamma_{12})^2 \Biggr\} + \frac{N_2}{3 \eta} \Biggl\{ \frac{(\gamma_{12})^3}{Q_2} - \frac{2(\gamma_{12})^\frac{5}{2}}{Q_1} + \frac{Q_2}{{Q_1}^2} (\gamma_{12})^2 \Biggr\} \\ & \quad  + \frac{5}{21 \eta} \Biggl\{ \frac{N_2}{Q_2} (\gamma_{12})^3 + \frac{\overline{N_2} \overline{Q_2}}{{Q_2}^2}(\gamma_{12})^3 \left(\frac{\beta_{1\overline{2}}}{\gamma_{12}}\right)  - \frac{\overline{N_2}}{Q_2} (\beta_{2\overline{2}})^\frac{3}{2} (\gamma_{12})^3 - \frac{N_2 Q_2}{{Q_1}^2} (\gamma_{12})^2 - \frac{\overline{N_2} \overline{Q_2}}{{Q_1}^2} (\gamma_{12})^2 \left(\frac{\beta_{1\overline{2}}}{\gamma_{12}}\right)  \\ & \quad + \left. \frac{\overline{N_2} Q_2}{{Q_1}^2} (\gamma_{12})^2 (\beta_{2\overline{2}})^\frac{3}{2} \Biggr\} \Biggr] \frac{df_2(x)}{dx} \right|_{x_2} + \Biggl[ \frac{N_1}{3 \eta Q_1} - \frac{N_2 Q_2}{3 \eta {Q_1}^2} \gamma_{12} + \frac{N_{2}}{3\eta Q_{1}} (\gamma_{12})^\frac{3}{2}  + \frac{\overline{N_2}}{3 \eta Q_1} \Biggl\{ (\beta_{1\overline{2}})^\frac{3}{2} \\ & \quad - (\gamma_{12})^\frac{3}{2} (\beta_{2\overline{2}})^\frac{3}{2} - \frac{\overline{Q_2}}{Q_1}  \left(\frac{\beta_{1\overline{2}}}{\gamma_{12}}\right) \gamma_{12}  + \frac{Q_2}{Q_1} (\beta_{2\overline{2}})^\frac{3}{2} \gamma_{12} \Biggr\}  + \frac{5}{21 \eta} \Biggl\{ \frac{N_1}{Q_1} + \frac{N_2 Q_2}{{Q_1}^2} \gamma_{12}  - \frac{N_2}{Q_1} (\gamma_{12})^\frac{3}{2} \\ & \quad + \frac{\overline{N_2}}{Q_1} (\gamma_{12})^\frac{3}{2} (\beta_{2\overline{2}})^\frac{3}{2} - \frac{\overline{N_2}}{Q_1} (\beta_{1\overline{2}})^\frac{3}{2} -  \frac{\overline{N_2} Q_2}{{Q_1}^2} (\beta_{2\overline{2}})^\frac{3}{2} \gamma_{12} + \frac{\overline{N_2} \overline{Q_2}}{{Q_1}^2} \gamma_{12}  \left(\frac{\beta_{1\overline{2}}}{\gamma_{12}}\right) \Biggr\} \Biggr] \left. \frac{df_1(x)}{dx}\right|_{x_1}  
          \end{split}
          \label{105}
      \end{equation} 
\begin{equation}
                \begin{split}
                          K_{22} &= \frac{4}{7} \biggl\{ \frac{\left(\frac{\beta_{1\overline{2}}}{\gamma_{12}}\right)^\frac{5}{2}}{Q_{2}} - \frac{\left(\beta_{2\overline{2}
                          }\right)^{3}}{\overline{Q_{2}}} \left. \biggr\} \sqrt{x}  (\overline{f_{2}(x)})^\frac{5}{2} \right|_{\overline{x_{2}}} - \left. \frac{4}{7 Q_{2}} \sqrt{x} (f_{2}(x))^\frac{5}{2} \right|_{x_{2}}+ \frac{\overline{N_{2}}}{3 \eta} \biggl\{ \left(\frac{\beta_{1\overline{2}}}{\gamma_{12}}\right)^{2} \frac{\overline{Q_{2}}}{Q_{2}^{2}} - \frac{(\beta_{2\overline{2}})^\frac{3}{2} \left(\frac{\beta_{1\overline{2}}}{\gamma_{12}}\right)}{Q_{2}} \\ & \quad + \frac{5}{7 \overline{Q_{2}}} (\beta_{2\overline{2}})^{3}  - \frac{5 \overline{Q_{2}}}{7 Q_{2}^{2}} \left(\frac{\beta_{1\overline{2}}}{\gamma_{12}}\right)^{2} \left. \biggr\} \frac{d\overline{f_{2}(x)}}{dx} \right|_{\overline{x_{2}}} + \biggl[ \frac{5 N_{2}}{21 \eta Q_{2}} + \beta_{2\overline{2}} \frac{\overline{N_{2}}}{3 \eta Q_{2}} \biggl\{ (\beta_{2\overline{2}})^\frac{1}{2} - \frac{5}{7} (\beta_{2\overline{2}})^\frac{1}{2} \biggr\} - \frac{\overline{N_{2}} \overline{Q_{2}}}{3 \eta Q_{2}^{2}} \\ & \quad \left(\frac{\beta_{1\overline{2}}}{\gamma_{12}}\right)   + \frac{5 \overline{N_{2}} \overline{Q_{2}}}{21 \eta Q_{2}^{2}} \left(\frac{\beta_{1\overline{2}}}{\gamma_{12}}\right) \left. \biggr] \frac{df_{2}(x)}{dx} \right|_{x_{2}} + (\beta_{2\overline{2}})^3\frac{\overline{N_2}}{3 \eta \overline{Q_2} } \left. \frac{d\overline{f_2(x)}}{dx} \right|_{\overline{x_2}} - \left. \left(\frac{\beta_{1\overline{2}}}{\gamma_{12}}\right) (\beta_{2\overline{2}})^\frac{3}{2}\frac{\overline{N_2}}{3 \eta Q_2} \frac{d \overline{f_2(x)}}{dx} \right|_{\overline{x_2}}  \\ & \quad + \left. \frac{1}{Q_2} \frac{N_2}{3\eta} \frac{df_2(x)}{dx} \right|_{x_2}
                          \end{split}
                          \label{54}
                      \end{equation}
          
          \begin{equation}
   K_{\overline{2}\overline{2}} = - \left. \frac{4}{7\overline{Q_{2}}} \sqrt{x} (\overline{f_{2}(x)})^\frac{5}{2} \right|_{\overline{x_{2}}} + \frac{4 \overline{N_{2}}}{7 \eta \overline{Q_{2}}} \frac{d\overline{f_{2}(x)}}{dx}\biggr|_{\overline{x_{2}}}
\end{equation}
    \begin{equation}
       \begin{split}
           K_{12} &= K_{21} = \left.\frac{4}{7Q_2} \left(\frac{\beta_{1\overline{2}}}{\gamma_{12}}\right)^\frac{5}{2} (\gamma_{12})^\frac{3}{2} \sqrt{x} (\overline{f_2(x)})^\frac{5}{2} \right|_{\overline{x_2}} - \left. \frac{4}{7 \overline{Q_2}} (\beta_{1\overline{2}})^\frac{3}{2} (\beta_{2\overline{2}})^\frac{3}{2} \sqrt{x} (\overline{f_2(x)})^\frac{5}{2} \right|_{\overline{x_2}} - \frac{4}{7Q_2} (\gamma_{12})^\frac{3}{2} \sqrt{x} \\ & \quad \left. (f_2(x))^\frac{5}{2} \right|_{x_2}  + \left. \frac{5 \overline{N_2}}{21 \eta \overline{Q_2}} (\beta_{1\overline{2}})^\frac{3}{2} (\beta_{2\overline{2}})^\frac{3}{2} \frac{d\overline{f_2(x)}}{dx} \right|_{\overline{x_2}} - \left. \frac{\overline{N_2}}{3 \eta Q_2} (\gamma_{12})^\frac{3}{2} \left(\frac{\beta_{1\overline{2}}}{\gamma_{12}}\right) (\beta_{2\overline{2}})^\frac{3}{2} \frac{d\overline{f_2(x)}}{dx} \right|_{\overline{x_2}} - \frac{5 \overline{N_2} \overline{Q_2}}{21 \eta {Q_2}^2 } \\ & \quad \left(\frac{\beta_{1\overline{2}}}{\gamma_{12}}\right)^2 (\gamma_{12})^\frac{3}{2} \left. \frac{d\overline{f_2(x)}}{dx} \right|_{\overline{x_2}}  + \left. \frac{\overline{N_2} \overline{Q_2}}{3 \eta {Q_2}^2} (\gamma_{12})^\frac{3}{2} \left(\frac{\beta_{1\overline{2}}}{\gamma_{12}}\right)^2 \frac{d\overline{f_2(x)}}{dx} \right|_{\overline{x_2}} + \left. \frac{\overline{N_2}}{3 \eta \overline{Q_2}} (\beta_{1\overline{2}})^\frac{3}{2} (\beta_{2\overline{2}})^\frac{3}{2} \frac{d\overline{f_2(x)}}{dx} \right|_{\overline{x_2}} \\ & \quad - \left. \frac{\overline{N_2}}{3 \eta Q_2} (\beta_{1\overline{2}})^\frac{3}{2} \left(\frac{\beta_{1\overline{2}}}{\gamma_{12}}\right) \frac{d\overline{f_2(x)}}{dx} \right|_{\overline{x_2}}  + \left. \frac{5N_2}{21 \eta Q_2} (\gamma_{12})^\frac{3}{2} \frac{df_2(x)}{dx} \right|_{x_2}  + \left.  \frac{5 \overline{N_2} \overline{Q_2}}{21 \eta {Q_2}^2} (\gamma_{12})^\frac{3}{2} \left(\frac{\beta_{1\overline{2}}}{\gamma_{12}}\right) \frac{df_2(x)}{dx} \right|_{x_2} \\ & \quad - \left. \frac{5 \overline{N_2}}{21 \eta Q_2} (\beta_{2\overline{2}})^\frac{3}{2} (\gamma_{12})^\frac{3}{2} \frac{df_2(x)}{dx}\right|_{x_2}  - \left. \frac{\overline{N_2}\overline{Q_2}}{3 \eta {Q_2}^2} (\gamma_{12})^\frac{3}{2} \left(\frac{\beta_{1\overline{2}}}{\gamma_{12}}\right) \frac{df_2(x)}{dx} \right|_{x_2} - \left. \frac{N_2}{3 \eta Q_1} \gamma_{12} \frac{df_2(x)}{dx} \right|_{x_2}  \\ & \quad + \left. \frac{\overline{N_2}}{3 \eta Q_2} (\beta_{1\overline{2}})^\frac{3}{2} \frac{df_2(x)}{dx} \right|_{x_2} + \left. \frac{N_2}{3 \eta Q_2} (\gamma_{12})^\frac{3}{2} \frac{df_2(x)}{dx}\right|_{x_2} + \left. \frac{N_2}{3 \eta Q_1} \frac{df_1(x)}{dx} \right|_{x_1}
           \end{split}
           \end{equation}
\begin{equation}
    K_{\overline{2}2} = - \left. \frac{4}{7\overline{Q_2}}(\beta_{2\overline{2}})^\frac{3}{2}\sqrt{x}(\overline{f_2(x)})^\frac{5}{2}\right|_{\overline{x_2}}+\frac{\overline{N_2}}{3\eta} \biggl\{\frac{(\beta_{2\overline{2}})^\frac{3}{2}}{\overline{Q_2}}-\frac{\left(\frac{\beta_{1\overline{2}}}{\gamma_{12}}\right)}{Q_2} +\left. \frac{5(\beta_{2\overline{2}})^\frac{3}{2}}{7\overline{Q_2}}\biggr\}\frac{d\overline{f_2(x)}}{dx}\right|_{\overline{x_2}} + \left. \frac{\overline{N_2}}{3\eta Q_2} \frac{df_2(x)}{dx}\right|_{x_2}
    \label{43}
\end{equation}
                       \begin{equation}
        \begin{split}
           K_{\overline{2}1}  &= -  \left.\frac{4}{7\overline{Q_2}} (\beta_{1\overline{2}})^\frac{3}{2} \sqrt{x} (\overline{f_2(x)})^\frac{5}{2} \right|_{\overline{x_2}} + \frac{\overline{N_2}}{3 \eta} \Biggl\{ \frac{5}{7 \overline{Q_2}} (\beta_{1\overline{2}})^\frac{3}{2}+ \left. \frac{(\beta_{1\overline{2}})^\frac{3}{2}}{\overline{Q_2}} - \frac{(\gamma_{12})^\frac{3}{2} \left(\frac{\beta_{1\overline{2}}}{\gamma_{12}}\right)}{Q_{2}} \Biggr\} \frac{d\overline{f_2(x)}}{dx}\right|_{\overline{x_2}}  \\ & \quad +   \frac{\overline{N_2}}{3 \eta}  \Biggl\{\frac{(\gamma_{12})^\frac{3}{2}}{Q_2}  - \left. \frac{\gamma_{12}}{Q_1} \Biggr\} \frac{df_2(x)}{dx} \right|_{x_2}  + \left. \frac{\overline{N_2}}{3 \eta Q_1} \frac{df_1(x)}{dx} \right|_{x_1}
             \label{66}
             \end{split}
\end{equation}
 The appendices used to evaluate the final expression of potential energy for Case I are given in \autoref{appendix_I}.
      \section{Case II: $1 < \eta \leq 10$  }
     In Case II, all quarks are distributed in the inner region, anti-charm and light quarks in the middle region, and the light quarks in the outermost region. It is present in more than one, $\eta>1$, pockets of the pentaquark system\cite{MohanGiri}.
   \begin{figure}[H]
\centering
\includegraphics[width=0.51\linewidth]{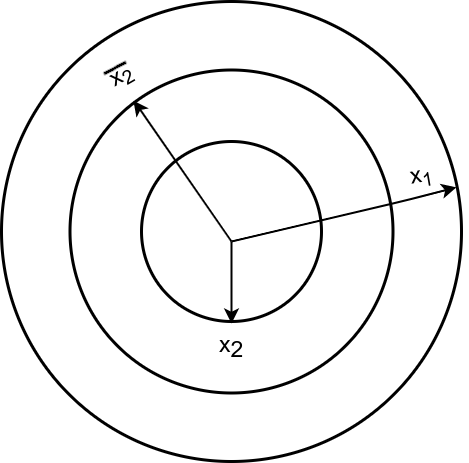} 
\caption{Distribution of quarks for the Case of $\eta = 2, 3, \cdots, 9, 10$}
\label{fig:2}
\end{figure}
As shown in fig \ref{fig:2}, in the innermost region, $0<x<x_2$, all three quarks are present. In the middle region, $x_{2}<x< \overline{x_{2}}$, anti-charm quarks and light quarks are present whereas in the outermost region, $\overline{x_{2}}<x< x_1$, only light quarks are present.
\subsection{Consistency Conditions}
Eqns \ref{yes}, \ref{no} \& \ref{dilemma} are consistency equations that govern the distribution of quarks in different regions of the pentaquark system.
\begin{itemize}
    \item \textbf{In region $0<x<x_{2}$}
\end{itemize}
In this region all three functions are present but we use the linear relationship of other functions with $f_{2}(x)$ as 
\begin{equation}
\begin{split}
    f_{1}(x) = \beta_{12}f_{2}(x) \\ 
    \overline{f_{2}(x)} = \beta_{\overline{2}2}f_{2}(x)
    \end{split}
    \label{bc_1}
\end{equation}
\hfill where, $\beta_{12}$ \& $\beta_{\overline{2}2}$ are proportionality constants \\ \\ 
Equation \ref{yes} gives
\begin{equation}
    \frac{d^{2}f_{2}(x)}{dx^{2}} = - \frac{6 \eta}{5 (5\eta -1)\overline{\alpha_{2}} \beta_{\overline{2}2}} \biggl\{ g_{1} (\beta_{12})^\frac{3}{2} + g_{2} \biggr\} \frac{(f_{2}(x))^\frac{3}{2}}{\sqrt{x}}
    \label{CC}
\end{equation}
Equation \ref{no} gives 
\begin{equation}
    \frac{d^{2}f_{2}(x)}{dx^{2}} = \left[- \frac{6 \eta}{5 (5 \eta -1) \beta_{12}} \overline{g_{2}} (\beta_{\overline{2}2})^\frac{3}{2} - \frac{9 \eta}{10 (5\eta-1) \beta_{12}} \left\{ \frac{(N_{1}g_{1} -1)}{N_{1}}(\beta_{12})^\frac{3}{2} + g_{2} \right\} \right] \frac{(f_{2}(x))^\frac{3}{2}}{\sqrt{x}}
    \label{CCR}
\end{equation}
Equation \ref{dilemma} gives
\begin{equation}
    \frac{d^{2}f_{2}(x)}{dx^{2}} = \left[ - \frac{6 \eta}{5 (5 \eta -1) \alpha_{2}} \overline{g_{2}} (\beta_{\overline{2}2})^\frac{3}{2} - \frac{9 \eta}{10 (5 \eta -1 ) \alpha_{2}} \biggl\{ \frac{(N_{2}g_{2} -1)}{N_{2}} + g_{1} (\beta_{12})^\frac{3}{2} \biggr\}\right] \frac{(f_{2}(x))^\frac{3}{2}}{\sqrt{x}}
    \label{CCRC}
\end{equation}
\\
In a single form eq$^{n}$s \ref{CC}, \ref{CCR}, \ref{CCRC} can be written as 
\begin{equation}
    \frac{d^{2}f_{2}(x)}{dx^{2}} = Q_{2} \frac{(f_{2}(x))^\frac{3}{2}}{\sqrt{x}}
    \label{com1}
\end{equation}
where, 
\begin{equation}
    Q_{2}  \begin{aligned}
        & = - \frac{6 \eta}{5 (5\eta -1)\overline{\alpha_{2}} \beta_{\overline{2}2}} \biggl\{ g_{1} (\beta_{12})^\frac{3}{2} + g_{2} \biggr\} \\ & 
        = - \frac{6 \eta}{5 (5 \eta -1) \beta_{12}} \overline{g_{2}} (\beta_{\overline{2}2})^\frac{3}{2} - \frac{9 \eta}{10 (5\eta-1) \beta_{12}} \left[ \frac{(N_{1}g_{1} -1)}{N_{1}}(\beta_{12})^\frac{3}{2} + g_{2} \right] \\ &
        = - \frac{6 \eta}{5 (5 \eta -1) \alpha_{2}} \overline{g_{2}} (\beta_{\overline{2}2})^\frac{3}{2} - \frac{9 \eta}{10 (5 \eta -1 ) \alpha_{2}} \biggl\{ \frac{(N_{2}g_{2} -1)}{N_{2}} + g_{1} (\beta_{12})^\frac{3}{2} \biggr\}
    \end{aligned}
    \label{con1}
\end{equation}\\
This second-order differential equation establishes the relationship between the charm function, $f_{2}(x)$, and the other functions present inside the region. 
\begin{itemize}
    \item \textbf{In region $x_{2}<x<\overline{x_{2}}$}
\end{itemize}
In this region, two functions $\overline{f_{2}(x)}$ \& $f_{1}(x)$ are present and we use linear relationship of $f_{1}(x)$ with $\overline{f_{2}(x)}$ as 
\begin{equation}
    f_{1}(x) = \gamma_{1\overline{2}} \overline{f_{2}(x)}
    \label{bc_2}
\end{equation}
Equation \ref{yes} gives 
\begin{equation}
    \frac{d^{2}\overline{f_{2}(x)}}{dx^{2}} = - \frac{6 \eta}{5 (5 \eta -1) \overline{\alpha_{2}}} g_{1} (\gamma_{1\overline{2}})^\frac{3}{2} \frac{(\overline{f_{2}(x)})^\frac{3}{2}}{\sqrt{x}}
    \label{d}
\end{equation}
Equation \ref{no} gives 
\begin{equation}
    \frac{d^{2}\overline{f_{2}(x)}}{dx^{2}} = \left[-\frac{6\eta}{5(5\eta-1) \gamma_{1\overline{2}}}    \overline{g_2} - \frac{9 \eta}{10 (5\eta -1)} \frac{(N_{1}g_{1} -1)}{N_{1}} (\gamma_{1\overline{2}})^\frac{1}{2}\right] \frac{(\overline{f_{2}(x)})^\frac{3}{2}}{\sqrt{x}}
    \label{e}
\end{equation}
Equation \ref{dilemma} gives 
\begin{equation*}
    0 = \frac{(\overline{f_{2}(x)})^\frac{3}{2}}{\sqrt{x}}
\end{equation*}
This leads to meaningless results. So we drop this equation. 
\\\\
In a single form eq$^{n}$s \ref{d}, \ref{e} can be written as 
\begin{equation}
    \frac{d^{2}\overline{f_{2}(x)}}{dx^{2}} = \overline{Q_{2}} \frac{(\overline{f_{2}(x)})^\frac{3}{2}}{\sqrt{x}}
    \label{com2}
\end{equation}
where, 
\begin{equation}
    \overline{Q_{2}}  \begin{aligned}
        & = - \frac{6 \eta}{5 (5 \eta -1) \overline{\alpha_{2}}} g_{1} (\gamma_{1\overline{2}})^\frac{3}{2}\\ & =  -\frac{6\eta}{5(5\eta-1) \gamma_{1\overline{2}}}    \overline{g_2} - \frac{9 \eta}{10 (5\eta -1)} \frac{(N_{1}g_{1} -1)}{N_{1}} (\gamma_{1\overline{2}})^\frac{1}{2}
    \end{aligned}
    \label{con2}
\end{equation}\\
 This second-order differential equation defines the link between the anti-charm function, $\overline{f_{2}(x)}$, and the other function present inside the region. 
             \begin{itemize}
   \item \textbf{In region $\overline{x_2}<x\leq x_1$}
    \end{itemize}
     In this region, $f_2(x) = 0$ \& $\overline{f_2(x)}=0$. 
    
    \noindent Equation \ref{yes} gives 
    \begin{equation}
           0 = -\frac{6\eta}{5(5\eta-1)} \biggl\{ g_1 \frac{(f_1(x))^\frac{3}{2}}{\sqrt{x}} \biggr\}
           \label{6}
    \end{equation}
    This leads to meaningless results. So we drop this equation. \\
    \noindent Equation \ref{no} gives 
   \begin{equation}
        \frac{d^2 f_1(x)}{dx^2}= -\frac{9\eta}{10(5\eta-1)}\left[\frac{(N_1 g_1-1)}{N_1} \frac{\left( f_1(x) \right)^\frac{3}{2}}{\sqrt{x}}\right]
        \label{time}
      \end{equation}
            \noindent Equation \ref{dilemma} gives 
      \begin{equation}
0 = - \frac{9\eta}{10(5\eta-1)}\left[g_1 \frac{\left( f_1(x) \right)^\frac{3}{2}}{\sqrt{x}} \right]
\label{8}
      \end{equation}
       This leads to meaningless results. So we drop this equation. \\
       
         In a single form eqn \ref{time} can be written as 
      \begin{equation}
    \frac{d^2 f_1(x)}{dx^2}= Q_1 \frac{\left( f_1(x) \right)^\frac{3}{2}}{\sqrt{x}}
    \label{com3}
      \end{equation}
     \begin{equation}\text{where, }Q_1 = \frac{-9\eta}{10(5\eta-1)}\frac{(N_1g_1 - 1)}{N_1}
     \label{con3}
       \end{equation}\\
       This second-order differential equation defines the existence of light quarks in the outermost region of the pentaquark system. \\ \\
        \noindent Therefore, eqns \ref{com1}, \ref{com2}, \ref{com3} are the required consistency conditions with the values of $Q_2$, $\overline{Q_2}$ \& $Q_1$ in eqns \ref{con1}, \ref{con2} \& \ref{con3} respectively. 
                \subsection{Expression of Kinetic Energy}
         After solving eqn \ref{23} and careful reduction, the final expression of kinetic energy for the Case II looks like
\begin{equation}
\begin{split}
    \frac{T}{C} &= \frac{4}{7} \biggl[ \alpha_{1}g_{1} \biggl\{ (\beta_{12})^\frac{5}{2} - (\gamma_{1\overline{2}})^\frac{5}{2} \left(\frac{\beta_{12}}{\gamma_{1\overline{2}}}\right)^\frac{5}{2} \biggr\} + \alpha_{2}g_{2} \left. \biggr] \sqrt{x} (f_{2}(x))^\frac{5}{2} \right|_{x_{2}} + \left. \frac{4}{7} \overline{\alpha_{2}} \overline{g_{2}} \sqrt{x} (\overline{f_{2}(x)})^\frac{5}{2} \right|_{\overline{x_{2}}} \\ & \quad + \left. \frac{4}{7} \alpha_{1} g_{1} \sqrt{x} (f_{1}(x))^\frac{5}{2} \right|_{x_{1}}  + \frac{5 N_{2}}{21 \eta}  \biggl[ \left(\frac{\beta_{12}}{\gamma_{1\overline{2}}}\right)^{2} \frac{Q_{2}}{\overline{Q_{2}}} \biggl\{ \alpha_{1} g_{1} (\gamma_{1\overline{2}})^\frac{5}{2} + \overline{\alpha_{2}} \overline{g_{2}} \biggr\} -\alpha_{1} g_{1} (\beta_{12})^\frac{5}{2} - \overline{\alpha_{2}} \overline{g_{2}} (\beta_{\overline{2}2})^\frac{5}{2} \\ & \quad - \alpha_{2} g_{2} \biggr] \left. \frac{df_{2}(x)}{dx} \right|_{x_{2}} + \frac{5}{21 \eta}  \biggl[ \frac{(\gamma_{1\overline{2}})^{2}}{Q_{1}} \alpha_{1}g_{1} \biggl\{ \overline{N_{2}} \overline{Q_{2}}  + \left(\frac{\beta_{12}}{\gamma_{1\overline{2}}}\right) N_{2}Q_{2} - (\beta_{\overline{2}2})^{\frac{3}{2}} N_{2}\overline{Q_{2}} \biggr\} - \biggl\{ \alpha_{1} g_{1} (\gamma_{1\overline{2}})^\frac{5}{2} + \overline{\alpha_{2}} \overline{g_{2}} \biggr\} \\ & \quad \biggl\{ \overline{N_{2}} + \left(\frac{\beta_{12}}{\gamma_{1\overline{2}}}\right) \frac{N_{2} Q_{2}}{\overline{Q_{2}}} - (\beta_{\overline{2}2})^\frac{3}{2} N_{2} \biggr\} \left. \biggr] \frac{d\overline{f_{2}(x)}}{dx} \right|_{\overline{x_{2}}} - \frac{5 \alpha_{1}g_{1}}{21 \eta} \biggl[ N_{1} + \gamma_{1\overline{2}} \frac{\overline{N_{2}} \overline{Q_{2}}}{Q_{1}} + (\gamma_{1\overline{2}})^{\frac{3}{2}} \biggl\{ (\beta_{\overline{2}2})^\frac{3}{2} N_{2} - \overline{N_{2}} \biggr\}  \\ & \quad - (\beta_{12})^\frac{3}{2} N_{2} - (\gamma_{1\overline{2}})(\beta_{\overline{2}2})^\frac{3}{2} \frac{N_{2} \overline{Q_{2}}}{Q_{1}} + \gamma_{1\overline{2}} \left(\frac{\beta_{12}}{\gamma_{1\overline{2}}}\right) \frac{N_{2} Q_{2}}{Q_{1}} \left. \biggr] \frac{df_{1}(x)}{dx} \right|_{x_{1}}
    \end{split}
\end{equation}
  The appendices used to evaluate the final expression of kinetic energy for Case II are given in \autoref{appendix_II}.
 \subsection{Expression of Potential Energy}
The seven different constants on which the potential energy of Case II depends are 
 \begin{equation}
      \begin{split}
         K_{11} & = \frac{4}{7} \biggl[ \frac{ \left(\frac{\beta_{12}}{\gamma_{1\overline{2}}}\right)^\frac{5}{2} (\gamma_{1\overline{2}})^3}{\overline{Q_2}} - \left. \frac{(\beta_{12})^3}{Q_{2}} \biggr] \sqrt{x} (f_{2}(x))^\frac{5}{2} \right|_{\overline{x_{2}}} + \left. \frac{4}{7} \biggl[\frac{(\gamma_{1\overline{2}})^\frac{5}{2}}{Q_1} - \frac{(\gamma_{1\overline{2}})^3}{\overline{Q_2}} \biggr] \sqrt{x} (\overline{f_{2}(x)})^\frac{5}{2} \right|_{x_2} \\ & \quad  -\left. \frac{4}{7 Q_1} \sqrt{x} (f_1(x))^\frac{5}{2} \right|_{x_1} + \Biggl[ \frac{N_{2}}{3 \eta} \Biggl\{ \frac{(\beta_{12})^3}{Q_{2}}  - \frac{(\beta_{12})^\frac{3}{2} (\gamma_{1\overline{2}})^\frac{3}{2}  \left(\frac{\beta_{12}}{\gamma_{1\overline{2}}}\right)}{\overline{Q_2}} + \frac{Q_{2}}{{\overline{Q_2}}^2} (\gamma_{1\overline{2}})^3  \left(\frac{\beta_{12}}{\gamma_{1\overline{2}}}\right)^2 \\ & \quad  - \frac{(\beta_{12})^\frac{3}{2} (\gamma_{1\overline{2}})^\frac{3}{2}  \left(\frac{\beta_{12}}{\gamma_{1\overline{2}}}\right)}{\overline{Q_2}} \Biggr\} + \frac{5 N_{2}}{21 \eta} \Biggl\{ \frac{(\beta_{12})^3}{Q_{2}} - \left. \frac{Q_{2}}{\overline{Q_2}^2}  \left(\frac{\beta_{12}}{\gamma_{1\overline{2}}}\right)^2 (\gamma_{1\overline{2}})^3 \Biggr\} \Biggr] \frac{df_{2}(x)}{dx} \right|_{x_{2}}  + \Biggl[ \frac{N_{2}}{3 \eta} \Biggl\{ \frac{(\beta_{12})^\frac{3}{2} (\gamma_{1\overline{2}})^\frac{3}{2}}{\overline{Q_2}}  \\ & \quad - \frac{(\beta_{12})^\frac{3}{2} \gamma_{1\overline{2}}}{Q_1} - \frac{(\beta_{\overline{2}2})^\frac{3}{2}(\gamma_{1\overline{2}})^3}{\overline{Q_2}} + \frac{(\beta_{\overline{2}2})^\frac{3}{2} (\gamma_{1\overline{2}})^\frac{5}{2}}{Q_1} + \frac{(\beta_{12})^\frac{3}{2} (\gamma_{1\overline{2}})^\frac{3}{2}}{\overline{Q_2}} - \frac{Q_{2}}{\overline{Q_2}^2} (\gamma_{1\overline{2}})^3  \left(\frac{\beta_{12}}{\gamma_{1\overline{2}}}\right) - \frac{(\beta_{12})^\frac{3}{2} \gamma_{1\overline{2}}}{Q_1}   \\ & \quad + \frac{(\gamma_{1\overline{2}})^\frac{5}{2} (\beta_{\overline{2}2})^\frac{3}{2}}{Q_1} + \frac{Q_{2}}{{Q_1}^2} (\gamma_{1\overline{2}})^2  \left(\frac{\beta_{12}}{\gamma_{1\overline{2}}}\right) - \frac{\overline{Q_2}}{{Q_1}^2} (\beta_{\overline{2}2})^\frac{3}{2} (\gamma_{1\overline{2}})^2 \Biggr\} + \frac{\overline{N_{2}}}{3 \eta} \Biggl\{ \frac{(\gamma_{1\overline{2}})^3}{\overline{Q_2}} - \frac{2(\gamma_{1\overline{2}})^\frac{5}{2}}{Q_1} + \frac{\overline{Q_2}}{{Q_1}^2} (\gamma_{1\overline{2}})^2 \Biggr\}  \\ & \quad + \frac{5}{21 \eta} \Biggl\{ \frac{\overline{N_{2}}}{\overline{Q_2}} (\gamma_{1\overline{2}})^3 + \frac{N_{2} Q_{2}}{\overline{Q_2}^2}(\gamma_{1\overline{2}})^3 \left(\frac{\beta_{12}}{\gamma_{1\overline{2}}}\right)  - \frac{N_{2}}{\overline{Q_2}} (\beta_{\overline{2}2})^\frac{3}{2} (\gamma_{1\overline{2}})^3 - \frac{\overline{N_{2}} \overline{Q_2}}{{Q_1}^2} (\gamma_{1\overline{2}})^2 - \frac{N_{2} Q_{2}}{{Q_1}^2} (\gamma_{1\overline{2}})^2 \left(\frac{\beta_{12}}{\gamma_{1\overline{2}}}\right)  \\ & \quad + \left. \frac{N_{2} \overline{Q_2}}{{Q_1}^2} (\gamma_{1\overline{2}})^2 (\beta_{\overline{2}2})^\frac{3}{2} \Biggr\} \Biggr] \frac{d\overline{f_{2}(x)}}{dx} \right|_{\overline{x_2}} + \Biggl[ \frac{N_1}{3 \eta Q_1}  - \frac{\overline{N_{2}} \overline{Q_2}}{3 \eta {Q_1}^2} \gamma_{1\overline{2}} + \frac{\overline{N_{2}}}{3\eta Q_{1}} (\gamma_{1\overline{2}})^\frac{3}{2}  + \frac{N_{2}}{3 \eta Q_1} \Biggl\{ (\beta_{12})^\frac{3}{2} \\ & \quad - (\gamma_{1\overline{2}})^\frac{3}{2} (\beta_{\overline{2}2})^\frac{3}{2}  - \frac{Q_{2}}{Q_1}  \left(\frac{\beta_{12}}{\gamma_{1\overline{2}}}\right) \gamma_{1\overline{2}}  + \frac{\overline{Q_2}}{Q_1} (\beta_{\overline{2}2})^\frac{3}{2} \gamma_{1\overline{2}} \Biggr\}  + \frac{5}{21 \eta} \Biggl\{ \frac{N_1}{Q_1} + \frac{\overline{N_{2}} \overline{Q_2}}{{Q_1}^2} \gamma_{1\overline{2}} - \frac{\overline{N_{2}}}{Q_1} (\gamma_{1\overline{2}})^\frac{3}{2} \\ & \quad + \frac{N_{2}}{Q_1} (\gamma_{1\overline{2}})^\frac{3}{2} (\beta_{\overline{2}2})^\frac{3}{2} - \frac{N_{2}}{Q_1} (\beta_{12})^\frac{3}{2} -  \frac{N_{2} \overline{Q_2}}{{Q_1}^2} (\beta_{\overline{2}2})^\frac{3}{2} \gamma_{1\overline{2}} + \frac{N_{2} Q_{2}}{{Q_1}^2} \gamma_{1\overline{2}}  \left(\frac{\beta_{12}}{\gamma_{1\overline{2}}}\right) \Biggr\} \Biggr] \left. \frac{df_1(x)}{dx}\right|_{x_1} 
          \end{split}
          \label{105}
      \end{equation} 
 \begin{equation}
   K_{22} = - \left. \frac{4}{7Q_{2}} \sqrt{x} (f_{2}(x))^\frac{5}{2} \right|_{x_{2}} + \frac{4 N_{2}}{7 \eta Q_{2}} \frac{df_{2}(x)}{dx}\biggr|_{x_{2}}
\end{equation}
\begin{equation}
                \begin{split}
                          K_{\overline{2}\overline{2}} &= \frac{4}{7} \biggl\{ \frac{\left(\frac{\beta_{12}}{\gamma_{1\overline{2}}}\right)^\frac{5}{2}}{\overline{Q_{2}}} - \frac{\left(\beta_{2\overline{2}
                          }\right)^{3}}{Q_{2}} \left. \biggr\} \sqrt{x}  (f_{2}(x))^\frac{5}{2} \right|_{x_{2}} - \left. \frac{4}{7 \overline{Q_{2}}} \sqrt{x} (\overline{f_{2}(x)})^\frac{5}{2} \right|_{\overline{x_2}}+ \frac{N_{2}}{3 \eta} \biggl\{ \left(\frac{\beta_{12}}{\gamma_{1\overline{2}}}\right)^{2} \frac{Q_{2}}{\overline{Q_{2}}^{2}} - \frac{(\beta_{\overline{2}2})^\frac{3}{2} \left(\frac{\beta_{12}}{\gamma_{1\overline{2}}}\right)}{\overline{Q_{2}}} \\ & \quad + \frac{5}{7 Q_{2}} (\beta_{\overline{2}2})^{3}  - \frac{5 Q_{2}}{7 \overline{Q_{2}}^{2}} \left(\frac{\beta_{12}}{\gamma_{1\overline{2}}}\right)^{2} \left. \biggr\} \frac{df_{2}(x)}{dx} \right|_{x_{2}} + \biggl[ \frac{5 \overline{N_{2}}}{21 \eta \overline{Q_{2}}} + \beta_{\overline{2}2} \frac{N_{2}}{3 \eta \overline{Q_{2}}} \biggl\{ (\beta_{\overline{2}2})^\frac{1}{2} - \frac{5}{7} (\beta_{\overline{2}2})^\frac{1}{2} \biggr\} - \frac{N_{2} Q_{2}}{3 \eta \overline{Q_{2}}^{2}} \\ & \quad \left(\frac{\beta_{12}}{\gamma_{1\overline{2}}}\right) + \frac{5 N_{2} Q_{2}}{21 \eta \overline{Q_{2}}^{2}} \left(\frac{\beta_{12}}{\gamma_{1\overline{2}}}\right) \left. \biggr] \frac{d\overline{f_{2}(x)}}{dx} \right|_{\overline{x_2}} + (\beta_{\overline{2}2})^3\frac{N_{2}}{3 \eta Q_{2} } \left. \frac{df_{2}(x)}{dx} \right|_{x_{2}}- \left. \left(\frac{\beta_{12}}{\gamma_{1\overline{2}}}\right) (\beta_{\overline{2}2})^\frac{3}{2}\frac{N_{2}}{3 \eta \overline{Q_2}} \frac{d f_{2}(x)}{dx} \right|_{x_{2}} \\ & \quad + \left. \frac{\overline{N_{2}}}{3\eta \overline{Q_2}} \frac{d\overline{f_{2}(x)}}{dx} \right|_{\overline{x_2}}
                          \end{split}
                          \label{54}
                      \end{equation}
                     \begin{equation}
        \begin{split}
           K_{21} & = K_{12}= -  \left.\frac{4}{7Q_{2}} (\beta_{12})^\frac{3}{2} \sqrt{x} (f_{2}(x))^\frac{5}{2} \right|_{x_{2}} + \frac{N_{2}}{3 \eta} \Biggl\{ \frac{5}{7 Q_{2}} (\beta_{12})^\frac{3}{2}+  \frac{(\beta_{12})^\frac{3}{2}}{Q_{2}} - \frac{(\gamma_{1\overline{2}})^\frac{3}{2} \left(\frac{\beta_{12}}{\gamma_{1\overline{2}}}\right)}{\overline{Q_{2}}} \Biggr\} \left. \frac{df_{2}(x)}{dx}\right|_{x_{2}} \\ & \quad +   \frac{N_{2}}{3 \eta}  \Biggl\{\frac{(\gamma_{1\overline{2}})^\frac{3}{2}}{\overline{Q_2}}  - \left. \frac{\gamma_{1\overline{2}}}{Q_1} \Biggr\} \frac{d\overline{f_{2}(x)}}{dx} \right|_{\overline{x_2}}  + \left. \frac{N_{2}}{3 \eta Q_1} \frac{df_1(x)}{dx} \right|_{x_1}
             \label{66}
             \end{split}
\end{equation}
\begin{equation}
    K_{\overline{2}2} = - \left. \frac{4}{7Q_{2}}(\beta_{\overline{2}2})^\frac{3}{2}\sqrt{x}(f_{2}(x))^\frac{5}{2}\right|_{x_{2}}+\frac{N_{2}}{3\eta} \biggl\{\frac{(\beta_{\overline{2}2})^\frac{3}{2}}{Q_{2}}-\frac{\left(\frac{\beta_{12}}{\gamma_{1\overline{2}}}\right)}{\overline{Q_2}} +\left. \frac{5(\beta_{\overline{2}2})^\frac{3}{2}}{7Q_{2}}\biggr\}\frac{df_{2}(x)}{dx}\right|_{x_{2}} + \left. \frac{N_{2}}{3\eta \overline{Q_2}} \frac{d\overline{f_{2}(x)}}{dx}\right|_{\overline{x_2}}
    \label{43}
\end{equation}
        \begin{equation}
       \begin{split}
           K_{\overline{2}1}&= \left.\frac{4}{7\overline{Q_2}} \left(\frac{\beta_{12}}{\gamma_{1\overline{2}}}\right)^\frac{5}{2} (\gamma_{1\overline{2}})^\frac{3}{2} \sqrt{x} (f_{2}(x))^\frac{5}{2} \right|_{x_{2}} - \left. \frac{4}{7 Q_{2}} (\beta_{12})^\frac{3}{2} (\beta_{\overline{2}2})^\frac{3}{2} \sqrt{x} (f_{2}(x))^\frac{5}{2} \right|_{x_{2}} - \frac{4}{7 \overline{Q_2}} (\gamma_{1\overline{2}})^\frac{3}{2} \sqrt{x} \\ & \quad \left. (\overline{f_{2}(x)})^\frac{5}{2} \right|_{\overline{x_2}}  + \left. \frac{5 N_{2}}{21 \eta Q_{2}} (\beta_{12})^\frac{3}{2} (\beta_{\overline{2}2})^\frac{3}{2} \frac{df_{2}(x)}{dx} \right|_{x_{2}} - \left. \frac{N_{2}}{3 \eta \overline{Q_2}} (\gamma_{1\overline{2}})^\frac{3}{2} \left(\frac{\beta_{12}}{\gamma_{1\overline{2}}}\right) (\beta_{\overline{2}2})^\frac{3}{2} \frac{df_{2}(x)}{dx} \right|_{x_{2}} -  \frac{5 N_{2} Q_{2}}{21 \eta \overline{Q_2}^2 } \\ & \quad \left(\frac{\beta_{12}}{\gamma_{1\overline{2}}}\right)^2 (\gamma_{1\overline{2}})^\frac{3}{2} \left. \frac{df_{2}(x)}{dx} \right|_{x_{2}}  + \left. \frac{N_{2} Q_{2}}{3 \eta \overline{Q_2}^2} (\gamma_{1\overline{2}})^\frac{3}{2} \left(\frac{\beta_{12}}{\gamma_{1\overline{2}}}\right)^2 \frac{df_{2}(x)}{dx} \right|_{x_{2}} + \left. \frac{N_{2}}{3 \eta Q_{2}} (\beta_{12})^\frac{3}{2} (\beta_{\overline{2}2})^\frac{3}{2} \frac{df_{2}(x)}{dx} \right|_{x_{2}} \\ & \quad - \left. \frac{N_{2}}{3 \eta \overline{Q_2}} (\beta_{12})^\frac{3}{2} \left(\frac{\beta_{12}}{\gamma_{1\overline{2}}}\right) \frac{df_{2}(x)}{dx} \right|_{x_{2}}  + \left. \frac{5\overline{N_{2}}}{21 \eta \overline{Q_2}} (\gamma_{1\overline{2}})^\frac{3}{2} \frac{d\overline{f_{2}(x)}}{dx} \right|_{\overline{x_2}}  + \left.  \frac{5 N_{2} Q_{2}}{21 \eta \overline{Q_2}^2} (\gamma_{1\overline{2}})^\frac{3}{2} \left(\frac{\beta_{12}}{\gamma_{1\overline{2}}}\right) \frac{d\overline{f_{2}(x)}}{dx} \right|_{\overline{x_2}} \\ & \quad - \left. \frac{5 N_{2}}{21 \eta \overline{Q_2}} (\beta_{\overline{2}2})^\frac{3}{2} (\gamma_{1\overline{2}})^\frac{3}{2} \frac{d\overline{f_{2}(x)}}{dx}\right|_{\overline{x_2}}  - \left. \frac{N_{2}Q_{2}}{3 \eta \overline{Q_2}^2} (\gamma_{1\overline{2}})^\frac{3}{2} \left(\frac{\beta_{12}}{\gamma_{1\overline{2}}}\right) \frac{d\overline{f_{2}(x)}}{dx} \right|_{\overline{x_2}} - \left. \frac{\overline{N_{2}}}{3 \eta Q_1} \gamma_{1\overline{2}} \frac{d\overline{f_{2}(x)}}{dx} \right|_{\overline{x_2}}  \\ & \quad + \left. \frac{N_{2}}{3 \eta \overline{Q_2}} (\beta_{12})^\frac{3}{2} \frac{d\overline{f_{2}(x)}}{dx} \right|_{\overline{x_2}} + \left. \frac{\overline{N_{2}}}{3 \eta \overline{Q_2}} (\gamma_{1\overline{2}})^\frac{3}{2} \frac{d\overline{f_{2}(x)}}{dx}\right|_{\overline{x_2}} + \left. \frac{\overline{N_{2}}}{3 \eta Q_1} \frac{df_1(x)}{dx} \right|_{x_1}
           \end{split}
           \end{equation}
           The appendices used to evaluate the final expression of potential energy for Case II are given in \autoref{appendix_II}.
        \section{Methods and Parameters}
      The TF quark model predicts the ground state energy, so the degeneracy factors $g_{I}$ and $\overline{g_{I}}$ are assigned with maximum possible values governed by the Pauli exclusion principle. \\   \\
        The mass of \textit{u} and \textit{d} quarks are very close to each other so we assume their density functions as the same. Each flavor, up and down, can get spin up and spin down giving rise to the degeneracy factor of 1 to 3 for light particles. Since charm and anti-charm quarks have different density functions, each of them can have a maximum degeneracy of 2. The different number of quarks and degeneracy factors that are used in this research is presented in the table \ref{methods and parameters} below. 
        \begin{table}[H]
            \centering
                        \caption{Number of quarks and their degeneracy factors}
                        \vspace{0.2cm}
             \begin{tabular}{|c c c c c c c c c c c c c|}
             \hline
             $\eta$ & Number of quarks & $N_{1}$ & & $g_{1}$ & & $N_{2}$ & & $g_{2}$ & & $\overline{N_{2}}$ & & $\overline{g_{2}}$\\
             \hline
              1 & 5  & 1 & &  3  & & 1 & & 1 & & 1  & & 1   \\
            \hline
              2 & 10  & 2 & &  3 & & 1 & & 2 & & 1  & & 2   \\
            \hline
              3 & 15  & 3 & & 3 & & 3 & & 1 & & 3  & & 1   \\
            \hline
              4 & 20  & 4 & & 3 & & 2 & & 2 & & 2  & & 2   \\
            \hline
              5 & 25  &  5 & & 3 & & 5 & & 1 & & 5  & & 1   \\
            \hline
              6 & 30  & 6 & & 3 & & 3 & & 2 & & 3  & & 2   \\
            \hline
               7 & 35  & 7 & & 3 & & 7 & & 1 & & 7  & & 1   \\
            \hline
               8 & 40  & 8 & & 3 & & 4 & & 2 & & 4  & & 2   \\
            \hline
               9 & 45  & 9 & & 3 & & 9 & & 1 & & 9  & & 1   \\
            \hline
               10 & 50  & 10 & & 3 & & 5 & & 2 & & 5  & & 2   \\
            \hline
            \end{tabular}
            \label{methods and parameters}
        \end{table}
        We used the mass of charm and anti-charm quarks from the paper \cite{SumanBaral}. The mass for charm and anti-charm quark was $1553$ MeV whereas the mass for light quark was $306$ MeV.   
          \section{Computational Details}
 The Mathematica program is used to evaluate mathematical equations related to the consistency conditions for different choices of pentaquarks. The second-order differential equations of eqns \ref{15}, \ref{21} \& \ref{9} are calculated using boundary conditions given in eqns \ref{11} \& \ref{17}. Similarly, the differential equations of eqns \ref{com1}, \ref{com2} \& \ref{com3} are calculated using boundary conditions given in eqns \ref{bc_1} \& \ref{bc_2}.\\\\
 The code begins by declaring and initializing several variables that are used throughout the program followed by different mathematical functions used to solve the program. These functions involve differential equations and boundary conditions that are solved numerically using the \texttt{NDSolve} function. The \texttt{NIntegrate} function is used to numerically integrate the normalization conditions and different integrals presented in the \autoref{appendix_I} and \autoref{appendix_II}. The \texttt{Evaluate} function is used to evaluate the density functions at the boundary between two regions. The calculation in the code is controlled by several loops and nested loops, which iteratively update the parameters and optimize the objective function. To minimize the energy, the code utilizes the \texttt{FindMinimum} function, employing an optimization algorithm such as gradient descent. This function effectively finds the local minimum of the objective function by iteratively adjusting the parameter values based on the gradient information until the prescribed convergence criterion is met. \\\\
 In addition to the optimization process, the \texttt{Plot} function of the Mathematica program plots graphs of the functions involved. The graphical outputs provide a comprehensive overview of the function landscape, facilitating further analysis and extraction of conclusions from the obtained results. Furthermore, XMGrace is used to plot graphs of kinetic, potential, volume, and total energy from the data extracted from the Mathematica program. These graphs help to draw conclusions from the research.  
\chapter{Results and Discussion}\label{Chapter 4}
\begin{figure}[H]
    \centering
    \includegraphics[width=17cm, height=10cm]{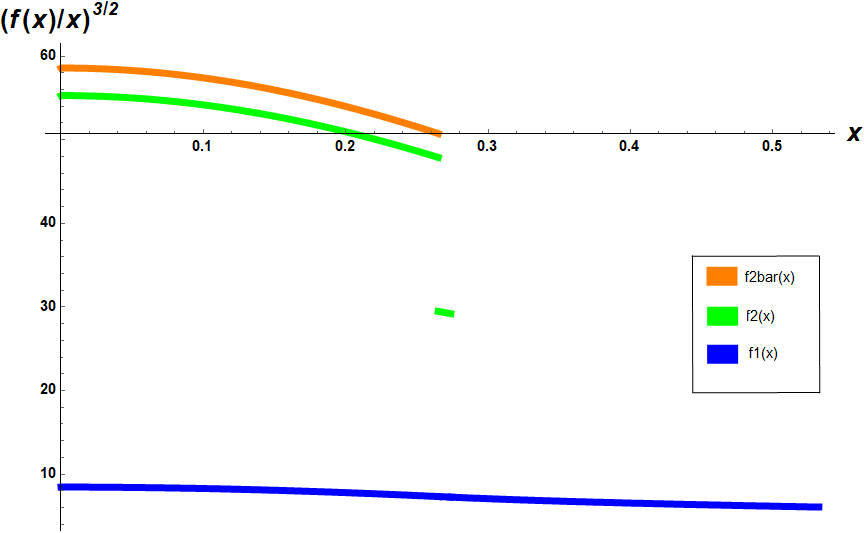}
    \caption{TF density functions for $\eta=1$}
    \label{all_1}
\end{figure}
Fig \ref{all_1} shows the density functions of quarks with respect to the dimensionless parameter \textit{x} for a single pocket of pentaquark, considered as Case I. It is seen that the density functions of pentaquarks decrease with the increase in \textit{x}. The anti-charm quark, depicted with the orange line, is confined inside the inner region only. The charm quark, denoted by the green line, shows a discontinuity at the boundary of the inner region. The discontinuity reveals that the density function of the charm quark decreases abruptly going on to the middle region. Furthermore, the blue curve denotes the light quark, though it has a very low density it radiates itself to the outermost boundary of the pentaquark system.
\begin{figure}[H]
\begin{minipage}{.5\textwidth}
\hspace{-1.5cm}\includegraphics[width=4in]{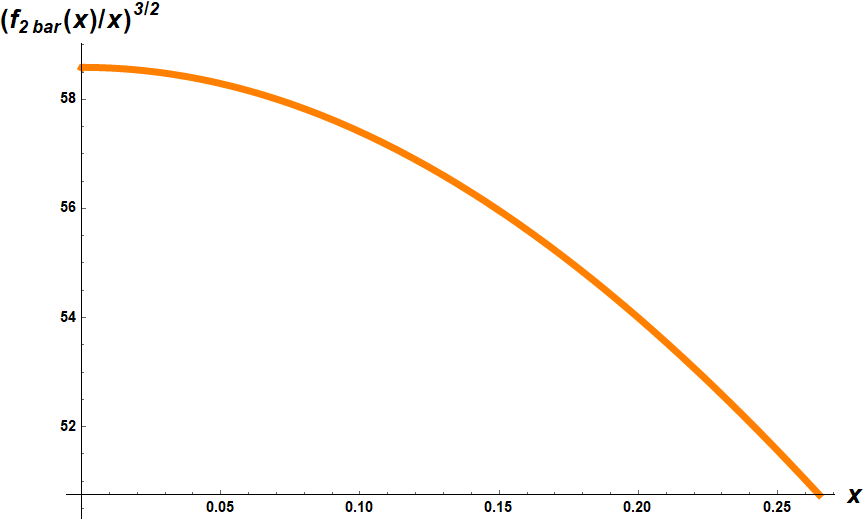}
       \hspace{-0.5cm} \caption{Density function for anti-charm quark}
\label{f2bar(x)}
\end{minipage}
\begin{minipage}{.5\textwidth}
\includegraphics[width=4in]{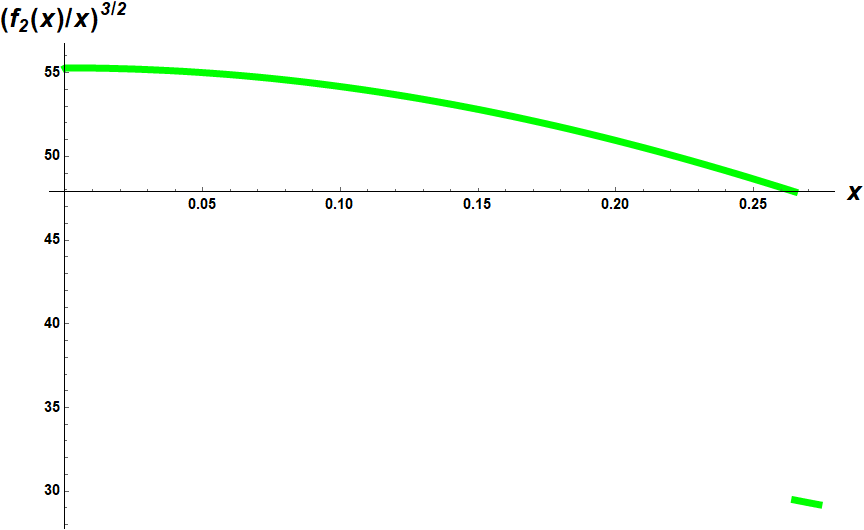}
         \caption{Density function for charm quark}
\label{f2(x)}
\end{minipage}
\end{figure} 

 \begin{wrapfigure}[8]{r}{0.5\textwidth}
 \vspace{-1cm}
\includegraphics[width=4in]{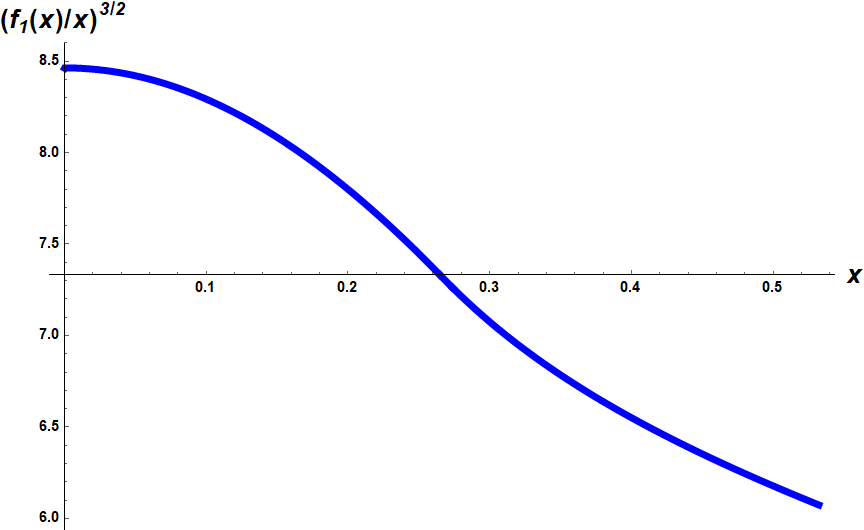} 
\caption{Density function for light quark}
\label{f1(x)}
\end{wrapfigure}
\begin{figure}[H]
    \centering
    \includegraphics[width=17cm, height=10cm]{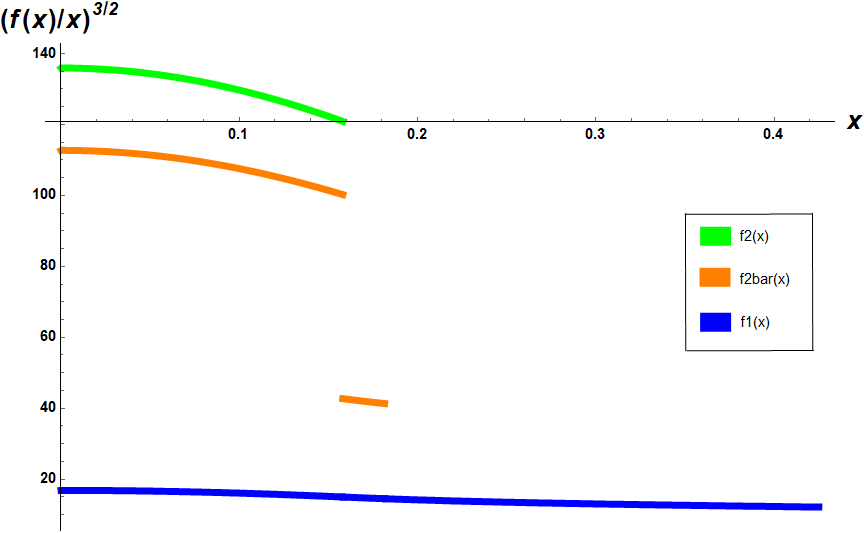}
    \caption{TF density functions for $\eta=2$}
    \label{all_2}
\end{figure}

Fig \ref{f2bar(x)}, \ref{f2(x)} and \ref{f1(x)} are the magnified image of density functions of anti-charm, charm, and light quarks respectively that are engraved in a single image of fig \ref{all_1}. The density function for the anti-charm quark decreases smoothly to the innermost boundary of the pentaquark whereas the density function for the charm quark has a clear discontinuity at the boundary of the inner region. \\ \\ 
Fig \ref{f1(x)} shows that the density function for light quarks is very low compared to other quarks and the function decreases smoothly to the outermost boundary of the pentaquark system. \\ \\ 
The nature of the plot of density functions for Case II is shown in fig \ref{all_2}.  The figure is actually for double pockets of pentaquarks but the nature of the graph is the same for all higher pockets of pentaquarks. The plots of TF density functions for higher pockets of pentaquarks are assembled in \autoref{appendix_III}. 
The difference in plots of density functions between the two cases is that $\overline{f_{2}(x)}$ lies in the inner region for Case I whereas for Case II $f_{2}(x)$ lies in that region. 
\begin{figure}[H]
    \centering
   \includegraphics[scale=0.70]{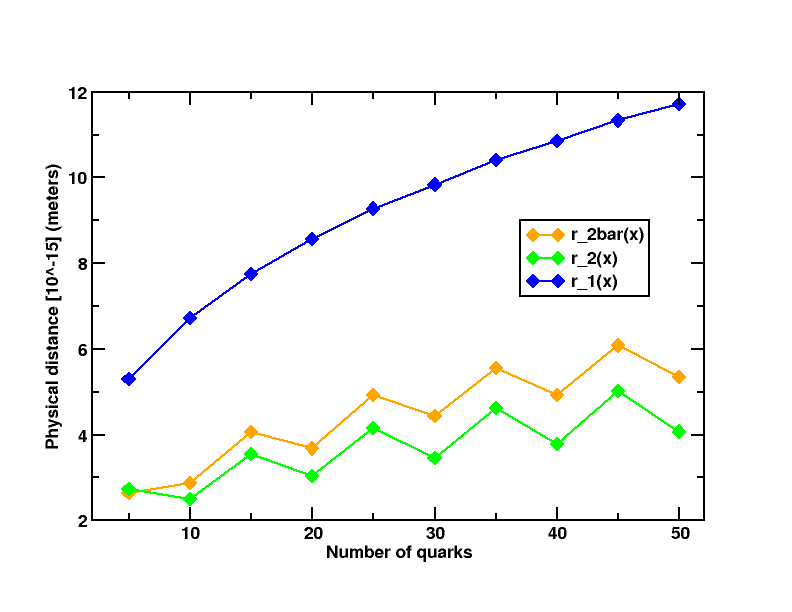}
    \caption{Physical distance versus quark content}
    \label{physicaldistance}
\end{figure}
In fig. \ref{physicaldistance}, the physical radius is plotted versus the quark number. The physical radius for the anti-charm quark is greater than the radius for the charm quark except at $\eta=1$ which is as expected. The physical distance is the product of constant $R$ and the dimensionless radius ${x}$. As the quark content increases, the dimensionless radius, ${x}$, becomes smaller whereas ${\eta^\frac{2}{3}}$ in ${R}$ increases. Therefore, the result is an increase in radius with increasing quark content. However, the curve of the radius plot tends to flatten out for a large number of quarks. \\
\begin{figure}[H]
    \centering
    \vspace{4cm}
   \includegraphics[width=20cm, height=15cm]{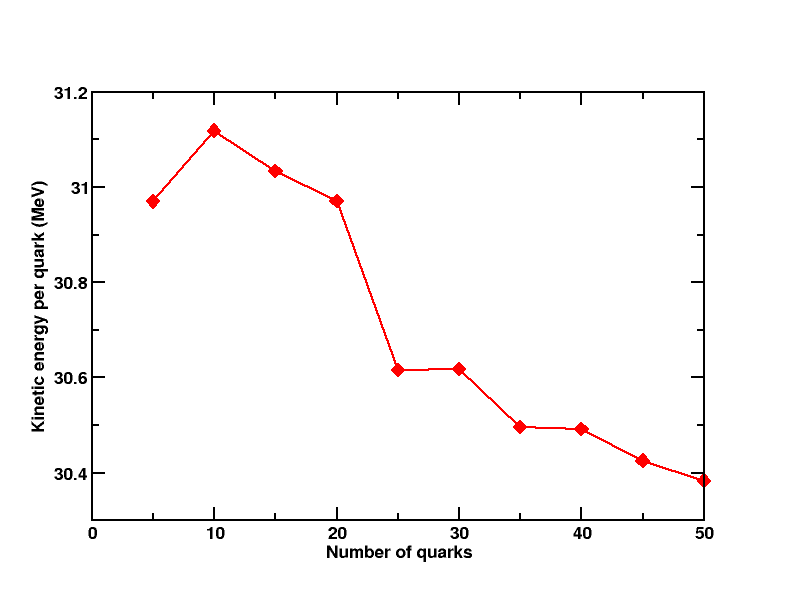}
    \caption{Kinetic energy per quark versus quark content}
    \label{kineticenergy}
\end{figure}
\begin{figure}[H]
    \centering
        \vspace{4cm}
   \includegraphics[width=20cm, height=15cm]{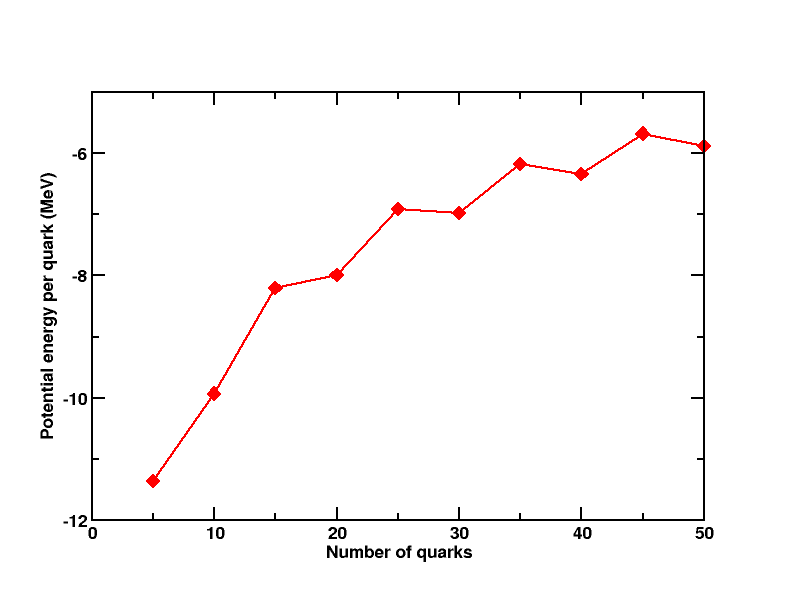}
    \caption{Potential energy per quark versus quark content}
    \label{potentialenergy}
\end{figure}
\begin{figure}[H]
    \centering
        \vspace{4cm}
   \includegraphics[width=20cm, height=15cm]{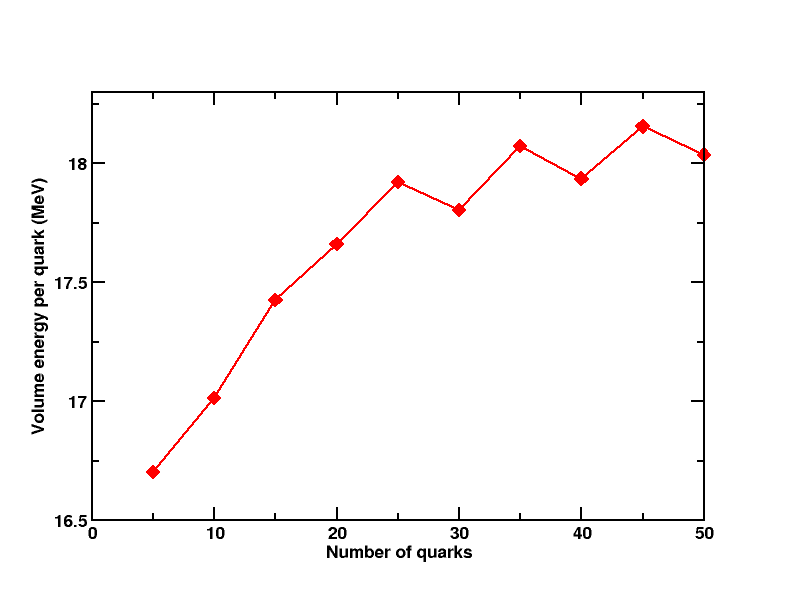}
    \caption{Volume energy per quark versus quark content}
    \label{volumeenergy}
\end{figure}
Within this model, we can identify three distinct forms of energy: kinetic energy, potential energy, and volume energy. As seen in fig. \ref{kineticenergy} the kinetic energy per quark decreases by a very small amount going from $\eta=1$ to $\eta=10$. Fig. \ref{potentialenergy} shows that the potential energy per quark increases smoothly up to $\eta=5$ but beyond that, the energy decreases and increases periodically. The nature of the graph for volume energy per quark is similar to that of potential energy per quark as seen in fig. \ref{volumeenergy}. We can observe dips in energies for even number of quarks and rise in energies for odd number of quarks beyond $\eta=5$. It is because the degeneracy is high for even number of quarks whereas it is low for odd number of quarks.
\vspace{-4.5cm}
\begin{figure}[H]
    \centering
        \vspace{4cm}
   \includegraphics[width=20cm, height=15cm]{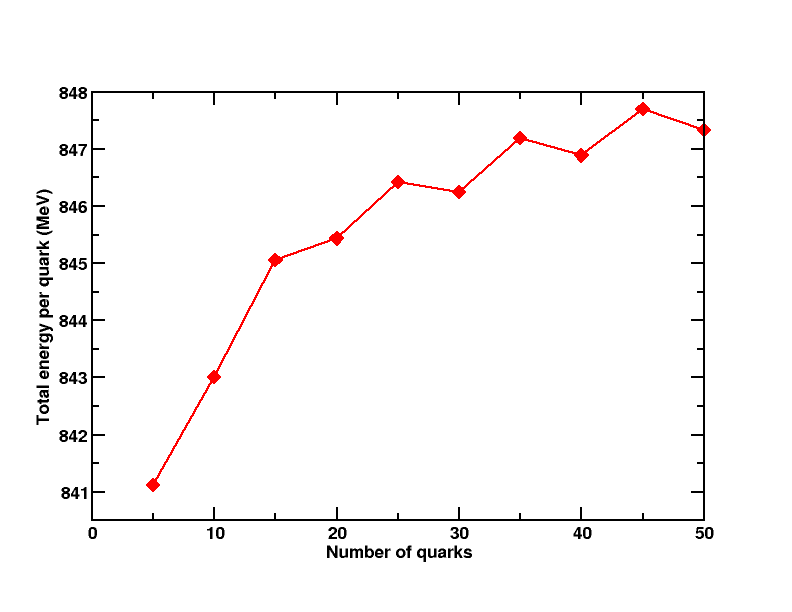}
    \caption{Total energy per quark versus quark content}
    \label{totalenergy}
\end{figure}
Fig. \ref{totalenergy} is the plot of total energy per quark i.e., the sum of kinetic, potential, and volume energy, against the quark content. It shows that the total energy per quark is minimum for a single pocket of pentaquark implying it is the most stable among families of pentaquarks. We can observe that the nature of the graph dominates over all energies except for kinetic energy. The degeneracy factor has played a significant role in this behavior of plots. 
\chapter{Conclusion}\label{Chapter 5}
We have studied the application of the TF quark model to the baryonic and mesonic pentaquark systems. \textit{$uudc\overline{c}$} is the pentaquark system that was under our study. We investigated the TF density functions at different boundaries of the pentaquark system. These boundaries were predicted by Mohan Giri \textit{et. al} in the paper \cite{MohanGiri}. The kinetic, potential, and volume energy were calculated for families of the pentaquark system. After calculating the total energy, a single pocket of pentaquark is considered the most stable because its energy was lowest compared to others.  \\ \\ 
In future research, one can incorporate spin interaction terms in the given pentaquark system. The study of the TF quark model can be extended to the decaquark system as well which can definitely assist in our goal of constructing a periodic table of exotic particles in the long run. 

\pagebreak
\chapter{Appendix I}
\label{appendix_I}
 \begin{equation}
        \int^{\overline{x_2}}_0 dx \sqrt{x} (\overline{f_2(x)})^\frac{3}{2} = \frac{\overline{N_2}}{3\eta}
    \end{equation}
     \begin{equation}
        \int^{x_2}_0 dx \sqrt{x} (f_2(x))^\frac{3}{2} = \frac{N_2}{3\eta}
    \end{equation}
     \begin{equation}
        \int^{x_1}_0 dx \sqrt{x} (f_1(x))^\frac{3}{2} = \frac{N_1}{3\eta}
    \end{equation}
      \begin{equation}
                   \int^{\overline{x_2}}_0 dx \sqrt{x} (f_2(x))^\frac{3}{2} = \frac{\overline{N_2}}{3 \eta} (\beta_{2\overline{2}})^\frac{3}{2}
                   \label{79}
                \end{equation}
                          \begin{equation}
     \int^{\overline{x_2}}_0 dx \sqrt{x} (f_1(x))^\frac{3}{2} = \frac{\overline{N_2}}{3 \eta} (\beta_{1\overline{2}})^\frac{3}{2}
     \label{90}
 \end{equation}
 \begin{equation}
   \int^{x_2}_{\overline{x_2}} dx \sqrt{x} (f_2(x))^\frac{3}{2}  = \frac{N_{2}}{3 \eta} - (\beta_{2\overline{2}})^\frac{3}{2}\frac{\overline{N_{2}}}{3 \eta}
    \label{59}
\end{equation}
    \begin{equation}
             \int^{x_2}_{\overline{x_2}} dx \sqrt{x} (f_2(x))^\frac{3}{2}  = \frac{1}{Q_2} \biggl[ x \frac{df_2(x)}{dx} - f_2(x) \biggr]_{x_2} - \frac{\left(\frac{\beta_{1\overline{2}}}{\gamma_{12}}\right)}{Q_2} \biggl[ x \frac{d\overline{f_2(x)}}{dx} - \overline{f_2(x)} \biggr]_{\overline{x_2}}
               \label{113}
           \end{equation}
           \begin{equation}
               \int^{x_1}_{x_2} dx \sqrt{x} (f_1(x))^\frac{3}{2}  = \frac{1}{Q_1} \biggl[ x \frac{df_1(x)}{dx} - f_1(x) \biggr]_{x_1} - \frac{\gamma_{12}}{Q_1} \biggl[ x \frac{df_2(x)}{dx} - f_2(x) \biggr]_{x_2}
               \label{114}
           \end{equation}
               \begin{equation}
            \biggl[ x \frac{d\overline{f_2(x)}}{dx}- \overline{f_2(x)} \biggr]_{\overline{x_2}} = \frac{\overline{N_2}\overline{Q_2}}{3\eta}
            \label{34}
        \end{equation}
        \begin{equation}
            \biggl[x\frac{df_2(x)}{dx}- f_2(x)\biggr]_{x_2} = \frac{N_2Q_2}{3\eta}+ \left(\frac{\beta_{1\overline{2}}}{\gamma_{12}}\right)\frac{\overline{N_2}\overline{Q_2}}{3\eta} - (\beta_{2\overline{2}})^\frac{3}{2} \frac{\overline{N_2}Q_2}{3\eta}
            \label{35}
        \end{equation}
            \begin{equation}
            \begin{split}
                \biggl[x\frac{df_1(x)}{dx}-f_1(x)\biggr]_{x_1} &= \frac{N_1Q_1}{3\eta}+\gamma_{12}\frac{N_2Q_2}{3\eta}-(\gamma_{12})^\frac{3}{2}\frac{N_2Q_1}{3\eta}+\biggl\{(\gamma_{12})^\frac{3}{2}(\beta_{2\overline{2}})^\frac{3}{2}\\ & \quad-(\beta_{1\overline{2}})^\frac{3}{2}\biggr\}\frac{\overline{N_2}Q_1}{3\eta} -\gamma_{12}(\beta_{2\overline{2}})^\frac{3}{2}\frac{\overline{N_2}Q_2}{3\eta}+\gamma_{12}\left(\frac{\beta_{1\overline{2}}}{\gamma_{12}}\right)\frac{\overline{N_2}\overline{Q_2}}{3\eta}
                \label{36}
                \end{split}
            \end{equation}
                             \begin{equation}
                          \int^{\overline{x_2}}_0 dx \frac{(\overline{f_2(x)})^\frac{5}{2}}{\sqrt{x}} = \left. \frac{4}{7} \sqrt{x} (\overline{f_2(x)})^\frac{5}{2} \right|_{\overline{x_2}} - \left. \frac{5\overline{N_2}}{21\eta}\frac{d\overline{f_2(x)}}{dx}\right|_{\overline{x_2}}
                          \label{38}
                      \end{equation}
                      \begin{equation}
                       \begin{split} \int^{x_2}_{\overline{x_2}} dx \frac{(f_2(x))^\frac{5}{2}}{\sqrt{x}} &= \left. \frac{4}{7} \sqrt{x} (f_2(x))^\frac{5}{2}\right|_{x_2} - \left. \frac{4}{7} \left(\frac{\beta_{1\overline{2}}}{\gamma_{12}}\right)^\frac{5}{2} \sqrt{x} (\overline{f_2(x)})^\frac{5}{2}\right|_{\overline{x_2}} -\left. \frac{5}{7Q_2}\frac{df_2(x)}{dx}\right|_{x_2}\biggl[\frac{N_2Q_2}{3\eta} \\ & \quad  + \left(\frac{\beta_{1\overline{2}}}{\gamma_{12}}\right) \frac{\overline{N_2}\overline{Q_2}}{3\eta} - (\beta_{2\overline{2}})^\frac{3}{2} \frac{\overline{N_2}Q_2}{3\eta}\biggr]  + \left. \frac{5}{7Q_2} \left(\frac{\beta_{1\overline{2}}}{\gamma_{12}}\right)^2\frac{d\overline{f_2(x)}}{dx}\right|_{\overline{x_2}}\frac{\overline{N_2}\overline{Q_2}}{3\eta}
                          \label{39}
                          \end{split}
                      \end{equation}
                      \begin{equation}                  
                      \begin{split}  
                      \int^{x_1}_{x_2}dx \frac{(f_1(x))^\frac{5}{2}}{\sqrt{x}} &=\left.\frac{4}{7}\sqrt{x}(f_1(x))^\frac{5}{2}\right|_{x_1} -\left. \frac{4}{7}(\gamma_{12})^\frac{5}{2} \sqrt{x} (f_2(x))^\frac{5}{2} \right|_{x_2}-\left.\frac{5}{7Q_1} \frac{df_1(x)}{dx}\right|_{x_1}\Biggl[\frac{N_1Q_1}{3\eta} \\ & \quad + \gamma_{12} \frac{N_2Q_2}{3\eta} - (\gamma_{12})^\frac{3}{2}\frac{N_2Q_1}{3\eta}   +\biggl\{(\gamma_{12})^\frac{3}{2}(\beta_{2\overline{2}})^\frac{3}{2} - (\beta_{1\overline{2}})^\frac{3}{2}\biggr\} \frac{\overline{N_2}Q_1}{3\eta} \\ & \quad - \gamma_{12}(\beta_{2\overline{2}})^\frac{3}{2}\frac{\overline{N_2}Q_2}{3\eta}   + \gamma_{12}\left(\frac{\beta_{1\overline{2}}}{\gamma_{12}}\right)\frac{\overline{N_2}\overline{Q_2}}{3\eta}\Biggr] +\left. \frac{5}{7Q_1}(\gamma_{12})^2
\frac{df_2(x)}{dx}\right|_{x_2} \biggl[ \frac{N_2Q_2}{3\eta}  \\ & \quad + \left(\frac{\beta_{1\overline{2}}}{\gamma_{12}}\right)\frac{\overline{N_2}\overline{Q_2}}{3\eta} - (\beta_{2\overline{2}})^\frac{3}{2} \frac{\overline{N_2}Q_2}{3\eta}\biggr]
\end{split}
\label{40}
\end{equation}
    \begin{equation}
    \int^{\overline{x_2}}_0 dx \frac{(\overline{f_2(x)})^\frac{3}{2}}{\sqrt{x}} =\left. \frac{1}{\overline{Q_2}} \frac{d\overline{f_2(x)}}{dx}\right|_{\overline{x_2}}
    \label{44}
\end{equation}
    \begin{equation}
    \text{In region $0<x\leq\overline{x_2}$,} \qquad 
      \int^{x}_0 dx \sqrt{x} (f_2(x))^\frac{3}{2}  =\frac{(\beta_{2\overline{2}})^\frac{3}{2}}{\overline{Q_2}}\biggl[x \frac{d\overline{f_2(x)}}{dx} - \overline{f_2(x)}\biggr]
        \label{46}
    \end{equation}
           \begin{equation}
                                   \int^{x}_{\overline{x_2}} dx \sqrt{x} (f_2(x))^\frac{3}{2}
     = \frac{1}{Q_2} \biggl[ x \frac{df_2(x)}{dx} - f_2(x) \biggr] - \frac{\overline{N_2}\overline{Q_2}}{3 \eta Q_2} \left(\frac{\beta_{1\overline{2}}}{\gamma_{12}}\right)
                    \label{80}
                \end{equation}
                 \begin{equation}
                  \text{Since $f_2(x)$ doesn't exist in region $> x_2$,} \qquad     \int^{x_1}_{x_2} dx \sqrt{x} (f_1(x))^\frac{3}{2} \int^{x_2}_{x} dx' \frac{(f_2(x'))^\frac{3}{2}}{\sqrt{x'}} = 0
                    \label{76}
                \end{equation}
           \begin{equation}
\begin{split}
  \text{In region $0<x\leq \overline{x_2}$,} \qquad & \int^{x_2}_{x} dx \frac{(f_2(x))^\frac{3}{2}}{\sqrt{x}}   = \left. \frac{(\beta_{2\overline{2}})^\frac{3}{2}}{\overline{Q_2}} \frac{d\overline{f_2(x)}}{dx}\right|_{\overline{x_2}} - \frac{(\beta_{2\overline{2}})^\frac{3}{2}}{\overline{Q_2}} \frac{d\overline{f_2(x)}}{dx} \\ & \quad + \left. \frac{1}{Q_2} \frac{df_2(x)}{dx}\right|_{x_2}  - \left. \frac{\left(\frac{\beta_{1\overline{2}}}{\gamma_{12}}\right)}{Q_2} \frac{d\overline{f_2(x)}}{dx}\right|_{\overline{x_2}}
    \label{57}
    \end{split}
\end{equation}
\begin{equation}
 \text{In region $\overline{x_2}<x \leq x_2$, \qquad
}\int^{x_2}_{x} dx \frac{(f_2(x))^\frac{3}{2}}{\sqrt{x}}   = \left.\frac{1}{Q_2} \frac{df_2(x)}{dx}\right|_{x_2} - \frac{1}{Q_2} \frac{df_2(x)}{dx}
    \label{58}
\end{equation}
                                             \begin{equation}
   \text{In region $0<x \leq \overline{x_2}$,} \qquad
 \int^{x}_0 dx \sqrt{x} (f_1(x))^\frac{3}{2}  = \frac{(\beta_{1\overline{2}})^\frac{3}{2}}{\overline{Q_2}} \Biggl[x \frac{d\overline{f_2(x)}}{dx} - \overline{f_2(x)}\Biggr]
    \label{67}
\end{equation}
                                \begin{equation}
                \begin{split}
                  \text{In region $\overline{x_2}<x\leq x_2$,}  \qquad &\int^{x}_0 dx \sqrt{x} (f_1(x))^\frac{3}{2}= \frac{\overline{N_2}}{3 \eta} (\beta_{1\overline{2}})^\frac{3}{2} - \frac{\overline{N_2} \overline{Q_2}}{3 \eta Q_2} (\gamma_{12})^\frac{3}{2} \left(\frac{\beta_{1\overline{2}}}{\gamma_{12}}\right) \\ & \quad + \frac{(\gamma_{12})^\frac{3}{2}}{Q_2} \biggl[ x \frac{df_2(x)}{dx} - f_2(x) \biggr]
                    \label{107}
                    \end{split}
                \end{equation}
                      \begin{equation}
      \begin{split}
         \text{In region $x_2<x\leq x_1$,} \qquad & \int^{x}_0 dx \sqrt{x} (f_1(x))^\frac{3}{2}  = \frac{\overline{N_2}}{3 \eta} (\beta_{1\overline{2}})^\frac{3}{2} + \frac{N_2}{3 \eta} (\gamma_{12})^\frac{3}{2} - \frac{N_2 Q_2}{3 \eta Q_1} \gamma_{12} \\ & \quad- \frac{\overline{N_2}}{3 \eta} (\gamma_{12})^\frac{3}{2} (\beta_{2\overline{2}})^\frac{3}{2}  - \frac{\overline{N_2} \overline{Q_2}}{3 \eta Q_1} \left(\frac{\beta_{1\overline{2}}}{\gamma_{12}}\right) \gamma_{12}  + \frac{\overline{N_2} Q_2}{3 \eta Q_1} (\beta_{2\overline{2}})^\frac{3}{2} \gamma_{12}\\ & \quad + \frac{1}{Q_1} \biggl[ x \frac{df_1(x)}{dx} - f_1(x) \biggr]
          \label{108}
          \end{split}
      \end{equation}
      \begin{equation}
      \begin{split}
      \text{In region $0<x\leq\overline{x_2}$,} \qquad  &\int^{x_1}_{x} dx \frac{(f_1(x))^\frac{3}{2}}{\sqrt{x}}   = \left.\frac{(\beta_{1\overline{2}})^\frac{3}{2}}{\overline{Q_2}} \biggl[ \frac{d\overline{f_2(x)}}{dx} \right|_{\overline{x_2}} - \frac{d\overline{f_2(x)}}{dx} \biggr] + \frac{(\gamma_{12})^\frac{3}{2}}{Q_2}  \left. \biggl[ \frac{df_2(x)}{dx} \right|_{x_2} \\ & \quad - \left. \left(\frac{\beta_{1\overline{2}}}{\gamma_{12}}\right) \frac{d\overline{f_2(x)}}{dx} \right|_{\overline{x_2}} \biggr] + \left. \frac{1}{Q_1} \biggl[ \frac{df_1(x)}{dx} \right|_{x_1} - \left.\gamma_{12} \frac{df_2(x)}{dx} \right|_{x_2} \biggr]
          \label{109}
          \end{split}
      \end{equation}
           \begin{equation}
           \begin{split}
             \text{In region $\overline{x_2}<x\leq x_2$,} \qquad & \int^{x_1}_{x} dx \frac{(f_1(x))^\frac{3}{2}}{\sqrt{x}}  = \left. \frac{(\gamma_{12})^\frac{3}{2}}{Q_2} \biggl[ \frac{df_2(x)}{dx} \right|_{x_2} - \frac{df_2(x)}{dx} \biggr] \\ & \quad + \left. \frac{1}{Q_1} \biggl[ \frac{df_1(x)}{dx} \right|_{x_1}  - \left.\gamma_{12} \frac{df_2(x)}{dx} \right|_{x_2} \biggr]
               \label{110}
               \end{split}
           \end{equation}
           \begin{equation}
             \text{In region $x_2<x\leq x_1$,} \qquad \int^{x_1}_{x} dx \frac{(f_1(x))^\frac{3}{2}}{\sqrt{x}} = \left.\frac{1}{Q_1} \biggl[ \frac{df_1(x)}{dx} \right|_{x_1} - \frac{df_1(x)}{dx} \biggr]
               \label{111}
                          \end{equation}
\chapter{Appendix II}
\label{appendix_II}
 \begin{equation}
        \int^{x_{2}}_0 dx \sqrt{x} (f_{2}(x))^\frac{3}{2} = \frac{N_{2}}{3\eta}
    \end{equation}
     \begin{equation}
        \int^{\overline{x_{2}}}_0 dx \sqrt{x} (\overline{f_{2}(x)})^\frac{3}{2} = \frac{\overline{N_2}}{3\eta}
    \end{equation}
     \begin{equation}
        \int^{x_1}_0 dx \sqrt{x} (f_1(x))^\frac{3}{2} = \frac{N_1}{3\eta}
    \end{equation}
      \begin{equation}
                   \int^{x_{2}}_0 dx \sqrt{x} (\overline{f_{2}(x)})^\frac{3}{2} = \frac{N_{2}}{3 \eta} (\beta_{\overline{2}2})^\frac{3}{2}
                   \label{79}
                \end{equation}
                          \begin{equation}
     \int^{x_{2}}_0 dx \sqrt{x} (f_1(x))^\frac{3}{2} = \frac{N_{2}}{3 \eta} (\beta_{12})^\frac{3}{2}
     \label{90}
 \end{equation}
 \begin{equation}
   \int^{\overline{x_{2}}}_{x_{2}} dx \sqrt{x} (\overline{f_{2}(x)})^\frac{3}{2}  = \frac{\overline{N_2}}{3 \eta} - (\beta_{\overline{2}2})^\frac{3}{2}\frac{N_{2}}{3 \eta}
    \label{59}
\end{equation}
    \begin{equation}
             \int^{\overline{x_{2}}}_{x_{2}} dx \sqrt{x} (\overline{f_{2}(x)})^\frac{3}{2}  = \frac{1}{\overline{Q_{2}}} \biggl[ x \frac{d\overline{f_{2}(x)}}{dx} - \overline{f_{2}(x)} \biggr]_{\overline{x_{2}}} - \frac{\left(\frac{\beta_{12}}{\gamma_{1\overline{2}}}\right)}{\overline{Q_{2}}} \biggl[ x \frac{df_{2}(x)}{dx} - f_{2}(x) \biggr]_{x_{2}}
               \label{113}
           \end{equation}
           \begin{equation}
               \int^{x_1}_{\overline{x_{2}}} dx \sqrt{x} (f_1(x))^\frac{3}{2}  = \frac{1}{Q_1} \biggl[ x \frac{df_1(x)}{dx} - f_1(x) \biggr]_{x_1} - \frac{\gamma_{1\overline{2}}}{Q_1} \biggl[ x \frac{d\overline{f_{2}(x)}}{dx} - \overline{f_{2}(x)} \biggr]_{\overline{x_{2}}}
               \label{114}
           \end{equation}
               \begin{equation}
            \biggl[ x \frac{df_{2}(x)}{dx}- f_{2}(x) \biggr]_{x_{2}} = \frac{N_{2}Q_{2}}{3\eta}
            \label{34}
        \end{equation}
        \begin{equation}
            \biggl[x\frac{d\overline{f_{2}(x)}}{dx}- \overline{f_{2}(x)}\biggr]_{\overline{x_{2}}} = \frac{\overline{N_2}\overline{Q_{2}}}{3\eta}+ \left(\frac{\beta_{12}}{\gamma_{1\overline{2}}}\right)\frac{N_{2}Q_{2}}{3\eta} - (\beta_{\overline{2}2})^\frac{3}{2} \frac{N_{2}\overline{Q_{2}}}{3\eta}
            \label{35}
        \end{equation}
            \begin{equation}
            \begin{split}
                \biggl[x\frac{df_1(x)}{dx}-f_1(x)\biggr]_{x_1} &= \frac{N_1Q_1}{3\eta}+\gamma_{1\overline{2}}\frac{\overline{N_2}\overline{Q_{2}}}{3\eta}-(\gamma_{1\overline{2}})^\frac{3}{2}\frac{\overline{N_2}Q_1}{3\eta}+\biggl\{(\gamma_{1\overline{2}})^\frac{3}{2}(\beta_{\overline{2}2})^\frac{3}{2}\\ & \quad-(\beta_{12})^\frac{3}{2}\biggr\}\frac{N_{2}Q_1}{3\eta} -\gamma_{1\overline{2}}(\beta_{\overline{2}2})^\frac{3}{2}\frac{N_{2}\overline{Q_{2}}}{3\eta}+\gamma_{1\overline{2}}\left(\frac{\beta_{12}}{\gamma_{1\overline{2}}}\right)\frac{N_{2}Q_{2}}{3\eta}
                \label{36}
                \end{split}
            \end{equation}
                             \begin{equation}
                          \int^{x_{2}}_0 dx \frac{(f_{2}(x))^\frac{5}{2}}{\sqrt{x}} = \left. \frac{4}{7} \sqrt{x} (f_{2}(x))^\frac{5}{2} \right|_{x_{2}} - \left. \frac{5N_{2}}{21\eta}\frac{df_{2}(x)}{dx}\right|_{x_{2}}
                          \label{38}
                      \end{equation}
                      \begin{equation}
                       \begin{split} \int^{\overline{x_{2}}}_{x_{2}} dx \frac{(\overline{f_{2}(x)})^\frac{5}{2}}{\sqrt{x}} &= \left. \frac{4}{7} \sqrt{x} (\overline{f_{2}(x)})^\frac{5}{2}\right|_{\overline{x_{2}}} - \left. \frac{4}{7} \left(\frac{\beta_{12}}{\gamma_{1\overline{2}}}\right)^\frac{5}{2} \sqrt{x} (f_{2}(x))^\frac{5}{2}\right|_{x_{2}} -\left. \frac{5}{7\overline{Q_{2}}}\frac{d\overline{f_{2}(x)}}{dx}\right|_{\overline{x_{2}}}\biggl[\frac{\overline{N_2}\overline{Q_{2}}}{3\eta} \\ & \quad  + \left(\frac{\beta_{12}}{\gamma_{1\overline{2}}}\right) \frac{N_{2}Q_{2}}{3\eta} - (\beta_{\overline{2}2})^\frac{3}{2} \frac{N_{2}\overline{Q_{2}}}{3\eta}\biggr]  + \left. \frac{5}{7\overline{Q_{2}}} \left(\frac{\beta_{12}}{\gamma_{1\overline{2}}}\right)^2\frac{df_{2}(x)}{dx}\right|_{x_{2}}\frac{N_{2}Q_{2}}{3\eta}
                          \label{39}
                          \end{split}
                      \end{equation}
                      \begin{equation}                  
                      \begin{split}  
                      \int^{x_1}_{\overline{x_{2}}}dx \frac{(f_1(x))^\frac{5}{2}}{\sqrt{x}} &=\left.\frac{4}{7}\sqrt{x}(f_1(x))^\frac{5}{2}\right|_{x_1} -\left. \frac{4}{7}(\gamma_{1\overline{2}})^\frac{5}{2} \sqrt{x} (\overline{f_{2}(x)})^\frac{5}{2} \right|_{\overline{x_{2}}}-\left.\frac{5}{7Q_1} \frac{df_1(x)}{dx}\right|_{x_1}\Biggl[\frac{N_1Q_1}{3\eta} \\ & \quad + \gamma_{1\overline{2}} \frac{\overline{N_2}\overline{Q_{2}}}{3\eta} - (\gamma_{1\overline{2}})^\frac{3}{2}\frac{\overline{N_2}Q_1}{3\eta}   +\biggl\{(\gamma_{1\overline{2}})^\frac{3}{2}(\beta_{\overline{2}2})^\frac{3}{2} - (\beta_{12})^\frac{3}{2}\biggr\} \frac{N_{2}Q_1}{3\eta} \\ & \quad - \gamma_{1\overline{2}}(\beta_{\overline{2}2})^\frac{3}{2}\frac{N_{2}\overline{Q_{2}}}{3\eta}   + \gamma_{1\overline{2}}\left(\frac{\beta_{12}}{\gamma_{1\overline{2}}}\right)\frac{N_{2}Q_{2}}{3\eta}\Biggr] +\left. \frac{5}{7Q_1}(\gamma_{1\overline{2}})^2
\frac{d\overline{f_{2}(x)}}{dx}\right|_{\overline{x_{2}}} \biggl[ \frac{\overline{N_2}\overline{Q_{2}}}{3\eta}  \\ & \quad + \left(\frac{\beta_{12}}{\gamma_{1\overline{2}}}\right)\frac{N_{2}Q_{2}}{3\eta} - (\beta_{\overline{2}2})^\frac{3}{2} \frac{N_{2}\overline{Q_{2}}}{3\eta}\biggr]
\end{split}
\label{40}
\end{equation}
    \begin{equation}
    \int^{x_{2}}_0 dx \frac{(f_{2}(x))^\frac{3}{2}}{\sqrt{x}} =\left. \frac{1}{Q_{2}} \frac{df_{2}(x)}{dx}\right|_{x_{2}}
    \label{44}
\end{equation}
    \begin{equation}
    \text{In region $0<x \leq x_{2}$,} \qquad 
      \int^{x}_0 dx \sqrt{x} (\overline{f_{2}(x)})^\frac{3}{2}  =\frac{(\beta_{\overline{2}2})^\frac{3}{2}}{Q_{2}}\biggl[x \frac{df_{2}(x)}{dx} - f_{2}(x)\biggr]
        \label{46}
    \end{equation}
           \begin{equation}
          \int^{x}_{x_{2}} dx \sqrt{x} (\overline{f_{2}(x)})^\frac{3}{2}
     = \frac{1}{\overline{Q_{2}}} \biggl[ x \frac{d\overline{f_{2}(x)}}{dx} - \overline{f_{2}(x)} \biggr] - \frac{N_{2}Q_{2}}{3 \eta \overline{Q_{2}}} \left(\frac{\beta_{12}}{\gamma_{1\overline{2}}}\right)
                    \label{80}
                \end{equation}
\begin{equation}
   \text{Since $f_{2}(x)$ doesn't exist in region $> x_{2}$,} \qquad  \int^{\overline{x_{2}}}_{x_{2}}dx \sqrt{x} (f_{1}(x))^\frac{3}{2} \int^{x_{2}}_{x} dx \frac{(f_{2}(x))^\frac{3}{2}}{\sqrt{x}} = 0 
\end{equation}
 \begin{equation}
\begin{split}
  \text{In region $0<x\leq x_{2}$,} \qquad \int^{\overline{x_{2}}}_{x} dx \frac{(\overline{f_{2}(x)})^\frac{3}{2}}{\sqrt{x}}  & = \left. \frac{(\beta_{\overline{2}2})^\frac{3}{2}}{Q_{2}} \frac{df_{2}(x)}{dx}\right|_{x_{2}} - \frac{(\beta_{\overline{2}2})^\frac{3}{2}}{Q_{2}} \frac{df_{2}(x)}{dx} \\ & \quad + \left. \frac{1}{\overline{Q_{2}}} \frac{d\overline{f_{2}(x)}}{dx}\right|_{\overline{x_{2}}}  - \left. \frac{\left(\frac{\beta_{12}}{\gamma_{1\overline{2}}}\right)}{\overline{Q_{2}}} \frac{df_{2}(x)}{dx}\right|_{x_{2}}
    \label{57}
    \end{split}
\end{equation}
\begin{equation}
 \text{In region $x_{2}<x \leq \overline{x_{2}}$, \qquad
}\int^{\overline{x_{2}}}_{x} dx \frac{(\overline{f_{2}(x)})^\frac{3}{2}}{\sqrt{x}}   = \left.\frac{1}{\overline{Q_{2}}} \frac{d\overline{f_{2}(x)}}{dx}\right|_{\overline{x_{2}}} - \frac{1}{\overline{Q_{2}}} \frac{d\overline{f_{2}(x)}}{dx}
    \label{58}
\end{equation}
       \begin{equation}
   \text{In region $0<x \leq x_{2}$,} \qquad
 \int^{x}_0 dx \sqrt{x} (f_1(x))^\frac{3}{2}  = \frac{(\beta_{12})^\frac{3}{2}}{Q_{2}} \Biggl[x \frac{df_{2}(x)}{dx} - f_{2}(x)\Biggr]
    \label{67}
\end{equation}
\begin{equation}
                \begin{split}
                  \text{In region $x_{2}<x\leq \overline{x_{2}}$,}  \qquad \int^{x}_0 dx \sqrt{x} (f_1(x))^\frac{3}{2}&= \frac{N_{2}}{3 \eta} (\beta_{12})^\frac{3}{2} - \frac{N_{2} Q_{2}}{3 \eta \overline{Q_{2}}} (\gamma_{1\overline{2}})^\frac{3}{2} \left(\frac{\beta_{12}}{\gamma_{1\overline{2}}}\right) \\ & \quad + \frac{(\gamma_{1\overline{2}})^\frac{3}{2}}{\overline{Q_{2}}} \biggl[ x \frac{d\overline{f_{2}(x)}}{dx} - \overline{f_{2}(x)} \biggr]
                    \label{107}
                    \end{split}
                \end{equation}
                      \begin{equation}
      \begin{split}
         \text{In region $\overline{x_{2}}<x\leq x_1$,} \qquad & \int^{x}_0 dx \sqrt{x} (f_1(x))^\frac{3}{2}  = \frac{N_{2}}{3 \eta} (\beta_{12})^\frac{3}{2} + \frac{\overline{N_2}}{3 \eta} (\gamma_{1\overline{2}})^\frac{3}{2} - \frac{\overline{N_2} \overline{Q_{2}}}{3 \eta Q_1} \gamma_{1\overline{2}} \\ & \quad- \frac{N_{2}}{3 \eta} (\gamma_{1\overline{2}})^\frac{3}{2} (\beta_{\overline{2}2})^\frac{3}{2}  - \frac{N_{2} Q_{2}}{3 \eta Q_1} \left(\frac{\beta_{12}}{\gamma_{1\overline{2}}}\right) \gamma_{1\overline{2}}  + \frac{N_{2} \overline{Q_{2}}}{3 \eta Q_1} (\beta_{\overline{2}2})^\frac{3}{2} \gamma_{1\overline{2}}\\ & \quad + \frac{1}{Q_1} \biggl[ x \frac{df_1(x)}{dx} - f_1(x) \biggr]
          \label{108}
          \end{split}
      \end{equation}
      \begin{equation}
      \begin{split}
      \text{In region $0<x \leq x_{2}$,} \qquad  &\int^{x_1}_{x} dx \frac{(f_1(x))^\frac{3}{2}}{\sqrt{x}}   = \left.\frac{(\beta_{12})^\frac{3}{2}}{Q_{2}} \biggl[ \frac{df_{2}(x)}{dx} \right|_{x_{2}} - \frac{df_{2}(x)}{dx} \biggr] + \frac{(\gamma_{1\overline{2}})^\frac{3}{2}}{\overline{Q_{2}}}  \left. \biggl[ \frac{d\overline{f_{2}(x)}}{dx} \right|_{\overline{x_{2}}} \\ & \quad - \left. \left(\frac{\beta_{12}}{\gamma_{1\overline{2}}}\right) \frac{df_{2}(x)}{dx} \right|_{x_{2}} \biggr] + \left. \frac{1}{Q_1} \biggl[ \frac{df_1(x)}{dx} \right|_{x_1} - \left.\gamma_{1\overline{2}} \frac{d\overline{f_{2}(x)}}{dx} \right|_{\overline{x_{2}}} \biggr]
          \label{109}
          \end{split}
      \end{equation}
           \begin{equation}
           \begin{split}
             \text{In region $x_{2}<x\leq \overline{x_{2}}$,} \qquad & \int^{x_1}_{x} dx \frac{(f_1(x))^\frac{3}{2}}{\sqrt{x}}  = \left. \frac{(\gamma_{1\overline{2}})^\frac{3}{2}}{\overline{Q_{2}}} \biggl[ \frac{d\overline{f_{2}(x)}}{dx} \right|_{\overline{x_{2}}} - \frac{d\overline{f_{2}(x)}}{dx} \biggr] \\ & \quad + \left. \frac{1}{Q_1} \biggl[ \frac{df_1(x)}{dx} \right|_{x_1}  - \left.\gamma_{1\overline{2}} \frac{d\overline{f_{2}(x)}}{dx} \right|_{\overline{x_{2}}} \biggr]
               \label{110}
               \end{split}
           \end{equation}
           \begin{equation}
             \text{In region $\overline{x_{2}}<x\leq x_1$,} \qquad \int^{x_1}_{x} dx \frac{(f_1(x))^\frac{3}{2}}{\sqrt{x}} = \left.\frac{1}{Q_1} \biggl[ \frac{df_1(x)}{dx} \right|_{x_1} - \frac{df_1(x)}{dx} \biggr]
               \label{111}
     \end{equation}
\chapter{Appendix III}
\label{appendix_III}

\begin{figure}[H]
    \centering
    \includegraphics[width=17cm, height=10cm]{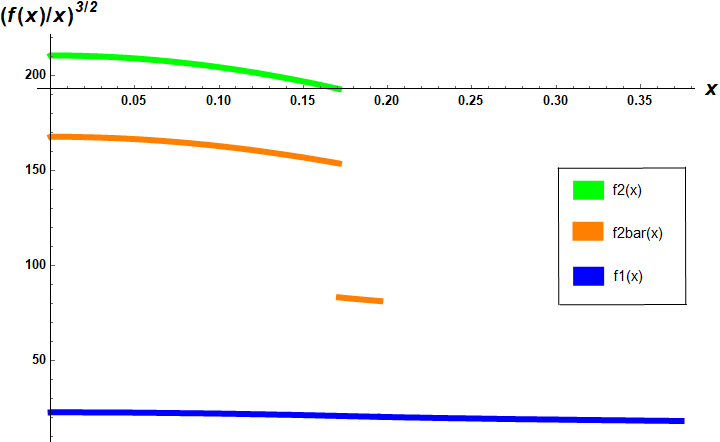}
    \caption{TF density functions for $\eta=3$}
    \label{all_3}
\end{figure}

\begin{figure}[H]
    \centering
    \includegraphics[width=17cm, height=10cm]{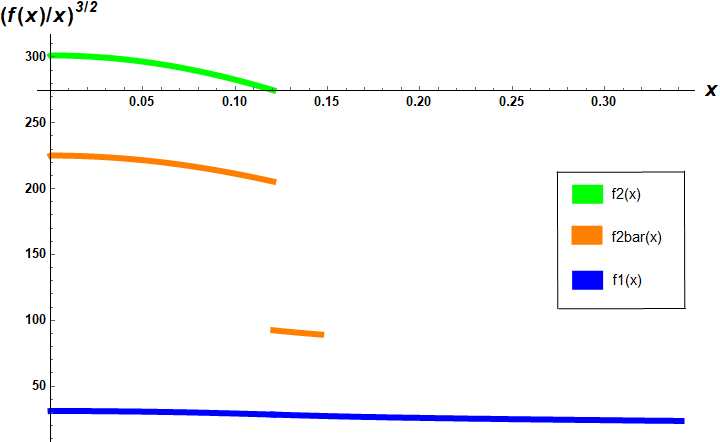}
    \caption{TF density functions for $\eta=4$}
    \label{all_4}
\end{figure}

\begin{figure}[H]
    \centering
    \includegraphics[width=17cm, height=10cm]{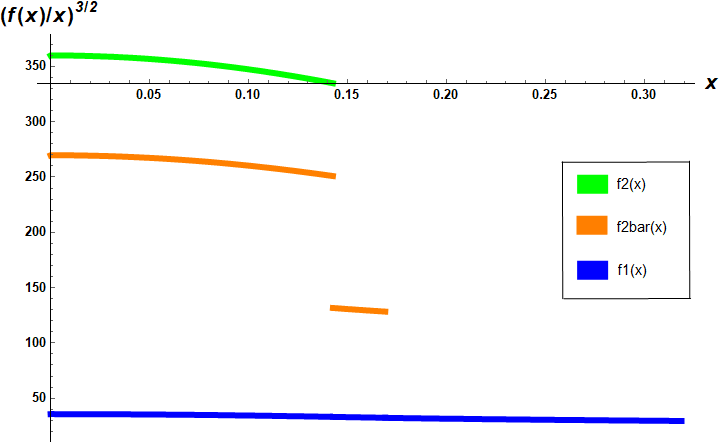}
    \caption{TF density functions for $\eta=5$}
    \label{all_5}
\end{figure}

\begin{figure}[H]
    \centering
    \includegraphics[width=17cm, height=10cm]{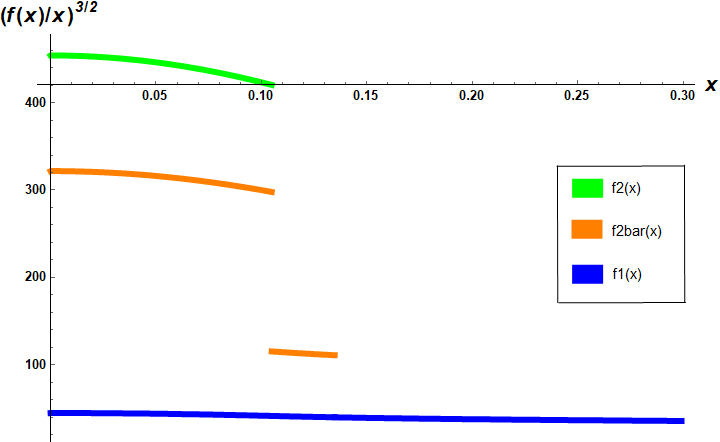}
    \caption{TF density functions for $\eta=6$}
    \label{all_6}
\end{figure}

\begin{figure}[H]
    \centering
    \includegraphics[width=17cm, height=10cm]{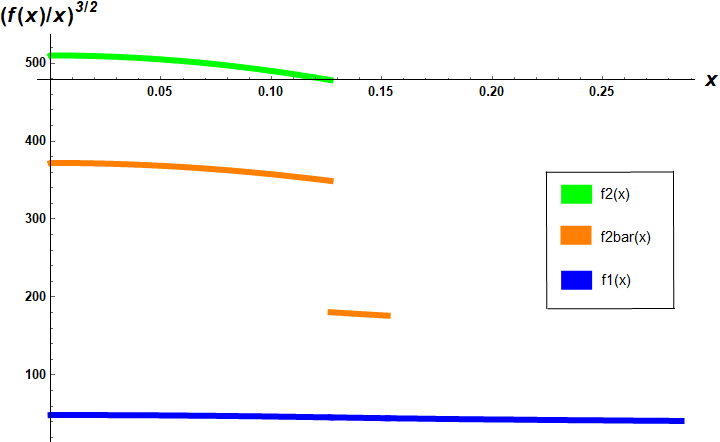}
    \caption{TF density functions for $\eta=7$}
    \label{all_7}
\end{figure}

\begin{figure}[H]
    \centering
    \includegraphics[width=17cm, height=10cm]{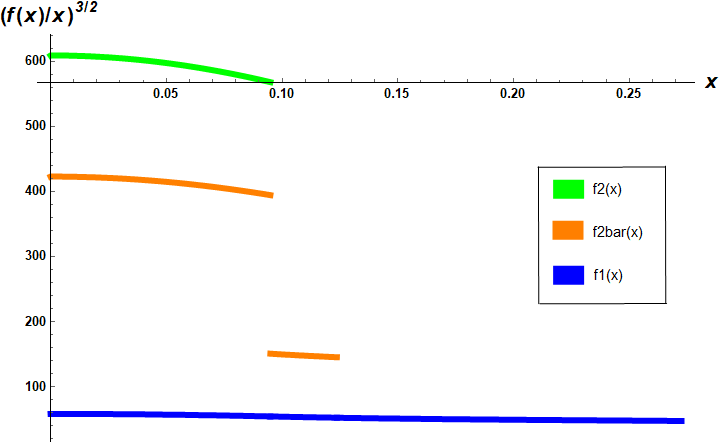}
    \caption{TF density functions for $\eta=8$}
    \label{all_8}
\end{figure}

\begin{figure}[H]
    \centering
    \includegraphics[width=17cm, height=10cm]{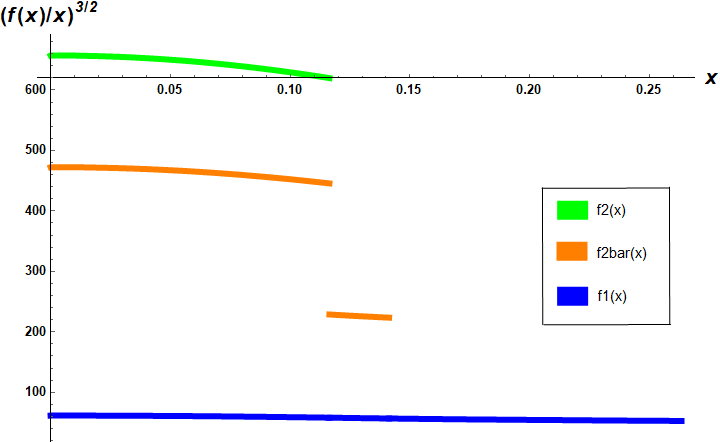}
    \caption{TF density functions for $\eta=9$}
    \label{all_9}
\end{figure}

\begin{figure}[H]
    \centering
    \includegraphics[width=17cm, height=10cm]{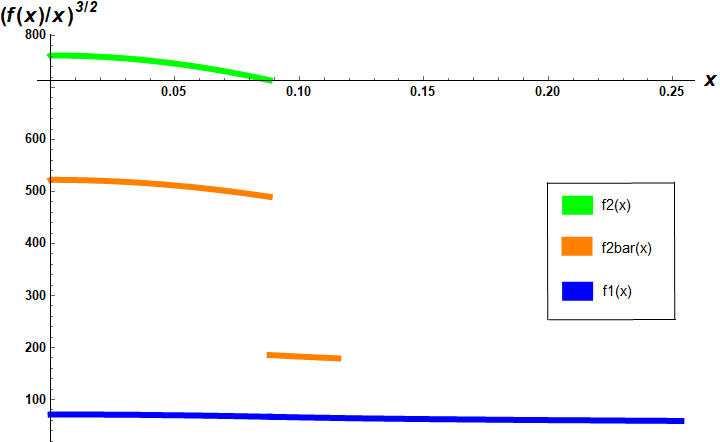}
    \caption{TF density functions for $\eta=10$}
    \label{all_10}
\end{figure}

\end{document}